\documentclass[11pt,preprint]{aastex}

\newcommand{\kms}{\,km~s$^{-1}$}

\newcommand{\Msun}{\mbox{\,$M_{\odot}$}}

\def\spose#1{\hbox to 0pt{#1\hss}}
\def\simlt{\mathrel{\spose{\lower 3pt\hbox{$\mathchar"218$}}
     \raise 2.0pt\hbox{$\mathchar"13C$}}}
\def\simgt{\mathrel{\spose{\lower 3pt\hbox{$\mathchar"218$}}
     \raise 2.0pt\hbox{$\mathchar"13E$}}}

\slugcomment{AJ in press}

\shorttitle{Keck/HST Study of Virgo Cluster Dwarfs}
\shortauthors{Geha et al.}

\begin{document}


\title{Internal Dynamics, Structure and Formation of Dwarf Elliptical
Galaxies: I.~A Keck/HST Study of Six Virgo Cluster Dwarfs}


\author{M.\ Geha}
\affil{UCO/Lick Observatory, University of California,
    Santa Cruz, 1156 High Street, Santa Cruz, CA 95064}
\email{mgeha@ucolick.org}

\author{P.\ Guhathakurta\altaffilmark{1,2}}
\affil{Herzberg Institute of Astrophysics, National Research Council
of Canada, 5071 West Saanich Road, Victoria, B.C., Canada V9E 2E7}
\email{raja@ucolick.org}
\altaffiltext{1}{Herzberg Fellow}
\altaffiltext{2}{Permanent address: UCO/Lick Observatory, University
    of California, Santa Cruz, 1156 High Street, Santa Cruz, CA 95064}

\and

\author{R.\ P.\ van der Marel}
\affil{Space Telescope Science Institute, 3700 San Martin Dr., Baltimore, MD 
21218}
\email{marel@stsci.edu}


\begin{abstract}
\renewcommand{\thefootnote}{\fnsymbol{footnote}}

Spectroscopy with the Keck~II 10-meter telescope\footnote{Data
presented herein were obtained at the W.\ M.\ Keck Observatory, which
is operated as a scientific partnership among the California Institute
of Technology, the University of California and the National
Aeronautics and Space Administration.  The Observatory was made
possible by the generous financial support of the W.\ M.\ Keck
Foundation.} and Echelle Spectrograph and Imager is presented for six
Virgo Cluster dwarf elliptical (dE) galaxies in the absolute magnitude
range $-15.7\le{M_V}\le-17.2$.  The mean line-of-sight velocity and
velocity dispersion are resolved as a function of radius along the
major axis of each galaxy, nearly doubling the total number of dEs
with spatially-resolved stellar kinematics.  None of the observed
objects shows evidence of strong rotation: upper limits on $v_{\rm
rot}/\sigma$, the ratio of the maximum rotational velocity to the mean
velocity dispersion, are well below those expected for
rotationally-flattened objects.  Such limits place strong constraints
on dE galaxy formation models.  Although these galaxies continue the
trend of low rotation velocities observed in Local Group dEs, they are
in contrast to recent observations of large rotation velocities in
slightly brighter cluster dEs.  Using surface photometry from {\it
Hubble Space Telescope\/}\footnote{Based on observations with the
NASA/ESA Hubble Space Telescope, obtained at the Space Telescope
Science Institute, which is operated by the Association of
Universities for Research in Astronomy, Inc., under NASA contract
NAS~5-26555.} Wide Field Planetary Camera~2 images and
spherically-symmetric dynamical models, we determine global
mass-to-light ratios $3\le\Upsilon_V\le6$.  These ratios are
comparable to those expected for an old to intermediate-age stellar
population and are broadly consistent with the observed $(V-I)$ colors
of the galaxies.  These dE galaxies therefore do not require a
significant dark matter component inside an effective radius.  We are
able to rule out central black holes more massive than
$\sim10^7\Msun$.  For the five nucleated dEs in our sample, kinematic
and photometric properties were determined for the central nucleus
separately from the underlying host dE galaxy.  These nuclei are as
bright or brighter than the most luminous Galactic globular clusters
and lie near the region of Fundamental Plane space occupied by
globular clusters.  In this space, the Virgo dE galaxies lie in the
same general region as Local Group and other nearby dEs, although
non-rotating dEs appear to have a slightly higher mean mass and
mass-to-light ratio than rotating dEs; the dE galaxies occupy a plane
parallel to, but offset from, that occupied by normal elliptical
galaxies.

\end{abstract}




\section{Introduction}\label{intro_sec}
\renewcommand{\thefootnote}{\fnsymbol{footnote}}

Dwarf elliptical galaxies (dEs) are the most common galaxy type by
number in the Local Universe, dominating the galaxy luminosity
function of nearby clusters.  Yet these galaxies remain among the most
poorly studied galaxies due to their faint luminosities, $M_V \ge
-18$, and characteristic low effective surface brightness $\mu_{V,\rm
eff}>22$~mag~arcsec$^{-2}$ \citep{fer94}.  Unlike brighter, classical
elliptical galaxies whose surface brightness profiles are well fit by
the de~Vaucouleurs $r^{1/4}$ law \citep{dev48}, dEs have brightness
profiles that are characterized by Sersic profiles \citep{ser68} with
indices ranging between $n=1$--3 (where $n=1$ corresponds to an
exponential law and $n=4$ to an $r^{1/4}$ law) making them appear more
diffuse than classical ellipticals of the same total magnitude
\citep{bin98}.  In the Virgo Cluster, the majority of dEs brighter
than $M_V \simlt -16$ contain compact central nuclei; fainter than
$M_V \simgt -12$ most dEs show no sign of a nucleus \citep{san85}.
Nuclei typically contain 5\% to 20\% of the total galaxy light and are
slightly resolved at the distance of Virgo by {\it Hubble Space
Telescope\/} ({\it HST\/}) imaging \citep{mil98}.

In hierarchical models of galaxy formation, dwarf galaxies form out of
small density fluctuations in the early Universe and are predicted to
be less spatially clustered than normal elliptical or spiral galaxies
\citep{dek86}.  However, dwarf elliptical galaxies are preferentially
found in dense cluster environments, more so than either ellipticals
or spirals \citep{bts87}; there are few, if any, examples of isolated
dEs.  Thus, current models favor dE formation from a progenitor galaxy
population.  The proposed progenitors of dEs are spiral or irregular
galaxies which are morphologically transformed into dEs through the
processes of galaxy harassment and interaction \citep*{moo98}.
Detailed internal kinematics of dEs are a powerful observational tool
with which to test these scenarios.

Until recently, radial velocity dispersion profiles were only
available for the Local Group dEs and two of the brightest dEs in the
Virgo Cluster \citep*{ben90,ben91,hel90}.  In addition, a handful of
global velocity dispersion measurements existed for dEs in various
environments outside the Local Group \citep{pet93}.  These
observations suggest that dEs have lower mass-to-light ratios than
Local Group dwarf spheroidals (e.g.,~Draco, Fornax) and are flattened
by velocity anisotropy rather than by rotation.  However,
\citet{ped02} and \citet{der01} recently presented kinematic profiles
for a few dEs in the Virgo and Fornax Clusters, respectively, with
rotation velocities comparable to that expected for a
rotationally-flattened spheroid.  These rotating dEs are more luminous
on average than the non-rotating dEs we have observed, and hint at a
possible association between dE luminosity and the presence of
rotation.

The question of whether dEs have significant rotation compared to
their velocity dispersion is particularly important in the context of
dE formation scenarios.  \citet{moo98} have demonstrated that the
process of galaxy harassment in cluster environments can
morphologically transform a spiral galaxy into a dE.  Although this
process tends to increase the velocity dispersion in a system, it is less
efficient at disrupting rotational motions and a significant fraction
of the progenitor's rotation is preserved.  Thus, measuring the amount
of internal angular momentum in dEs can constrain the progenitor
galaxy type and/or the amount of disruption required.

We present internal kinematics as a function of radius for a sample of
six~dE galaxies in the Virgo Cluster based on Keck observations.
These data are interpreted in conjunction with archival {\it HST\/}
imaging.  Preliminary results from this study were presented in
\citet*{geh02}.  This paper is organized as follows: in
\S\,\ref{data_sec} we present the Keck spectroscopic and {\it HST\/}
imaging observations of our target galaxies, along with an outline of
data reduction procedures; in \S\,\ref{res_sec} we present velocity
and velocity dispersion profiles and describe the dynamical models
that are applied to the data to derive mass-to-light ratios,
constraints on orbital anisotropy, and limits on the central black
hole mass; the broader implications of the results are discussed in
\S\,\ref{disc_sec}.

\section{The Data}\label{data_sec}

\begin{figure}
\plotone{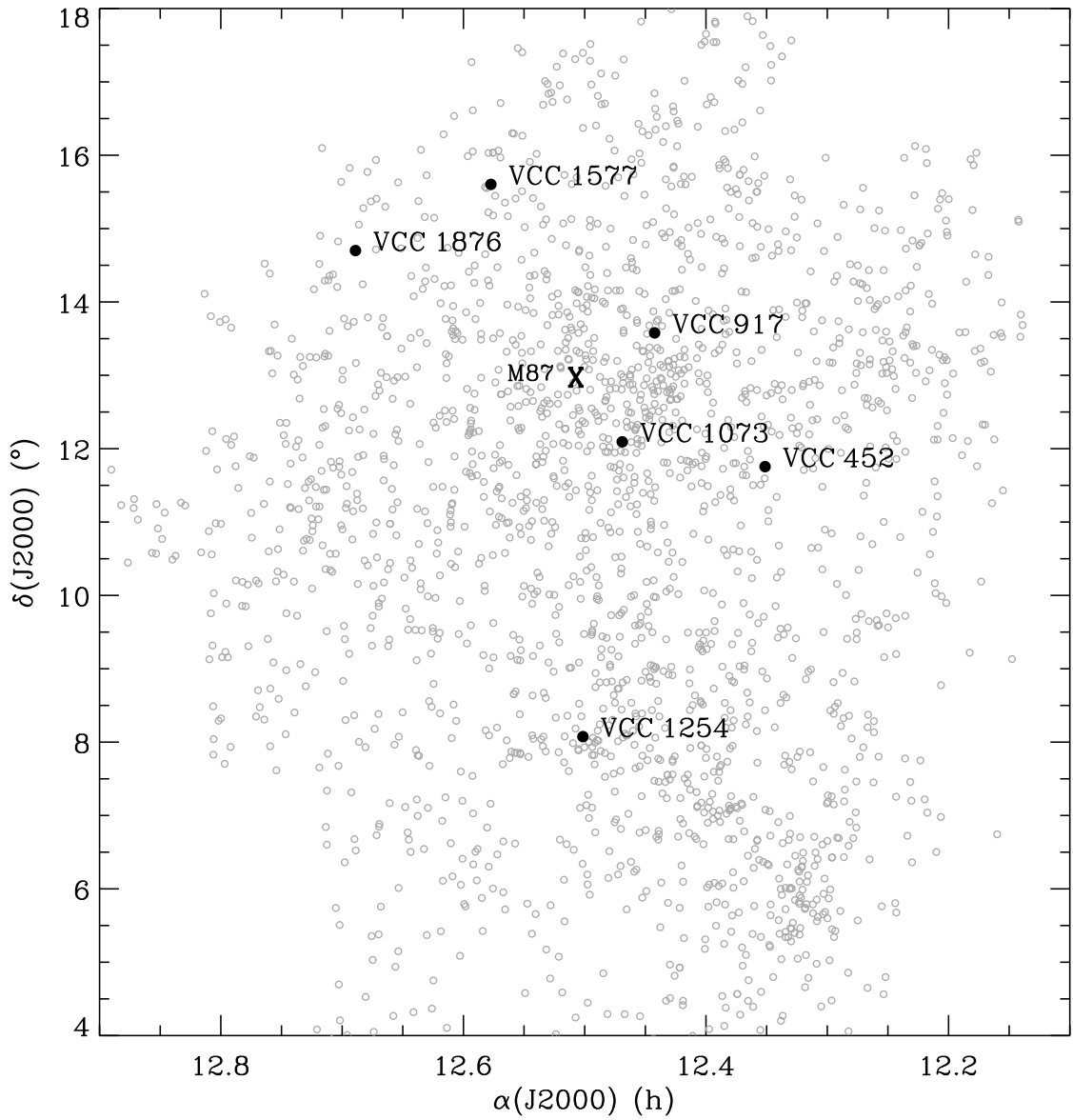}
\vskip 1 cm
\caption{Positions (J2000) of the six target Virgo Cluster dE galaxies
on the sky (solid circles) relative to all cluster members (open
circles) as identified by \citet{bin85}.  The ``X'' symbol indicates
the position of M87 at the center of the Virgo cluster.\label{vcc}}
\end{figure}

The Virgo Cluster is the closest large reservoir of dE galaxies beyond
the Local Group and presents a significantly different environment in
which to study such galaxies.  The six dE galaxies presented below
were drawn from the bright end of the Virgo dE luminosity function
and have been imaged with {\it HST\/}.  These objects lie at a variety of
distances from the center of the Virgo Cluster as shown in Figure~\ref{vcc}.
The positions and photometric properties of the observed galaxies are listed
in Table~1.  Foreground reddening values are taken
from \citet*{sch98} assuming a standard Galactic extinction law with
$R_V = 3.1$.  Throughout this paper a Virgo Cluster true distance modulus
of $(m - M)_0 = 30.92$ is adopted, i.e.,~a distance of 15.3~Mpc, as
determined by the {\it HST\/} Key Project on the extragalactic
distance scale \citep{fre01}.

\subsection{Spectroscopy}\label{spec_sec}
\subsubsection{Observations}\label{spec_obs_sec}

Six dE galaxies were observed on 2001 March 20--21 using the
Keck~II 10-m telescope and the Echelle Spectrograph and Imager
\citep[ESI;][]{she02}.  Observations were made in the echellette mode
with continuous wavelength coverage over the range
$\rm\lambda\lambda3900$--$11000\mbox{\AA}$ across 10~echelle orders with a
spectral dispersion of 11.4\kms~pixel$^{-1}$.  The spectra were
obtained through a $0.75'' \times 20''$ slit, resulting in an
instrumental resolution of 23\kms\ (Gaussian sigma) over the entire
spectrum, or $R \equiv (\lambda/\Delta\lambda) \approx 10,000$.  The
slit was positioned on the major axis of each galaxy, such that the galaxy's
center was displaced $\sim5''$ from the center of the slit along its $20''$
length.  Three consecutive exposures of 20~minutes were obtained for each
galaxy except VCC~1577, for which $5\times20$~minutes were obtained.
A summary of the observing parameters is given in Table~2.  High
signal-to-noise ratio spectra of giant stars covering the range of spectral
types G8III to M0III were taken with the same instrumental setup for
use as templates in the kinematic profile fitting described in
\S\,\ref{vp}.  Standards stars were observed both centered on and
trailed across the slit width in order to recover an accurate estimate
of the instrumental broadening for point and extended sources, respectively.

\subsubsection{Data Reduction}\label{datared_sec}

The ESI data were reduced using a combination of IRAF echelle and
long-slit spectral reduction tasks.  First, the overscan and a dark
frame, scaled by the exposure time, were subtracted from the data.
The sum of a bright calibration star spectral exposure and a flat
field spectral exposure were used to trace the ends of the slit as
well as a fiducial spatial point (the location of the bright star) for
each of the 10~curved echelle orders.  This process yielded an
empirical measurement of the spatial pixel scale for each order: it
varies from $0.13''$ in the bluest order to $0.18''$ in the reddest
order.  Scattered light was subtracted from individual frames by
fitting a smooth function to the areas outside these apertures (spaces
between echelle orders) using the APSCATTER task.  To preserve spatial
information, the APALL task was used in ``strip'' mode to extract and
rectify two-dimensional rectangular strips for each echelle order by
shifting and aligning each spatial column based on the aperture trace
information.  The rectified orders were then interpolated to a common
spatial pixel scale of $0.18''$ per pixel.  Calibration frames such
as flat field, arc lamp, and template star exposures were also
extracted into rectified, aligned and spatially-corrected strips using
the same procedure.

\begin{figure}
\plotone{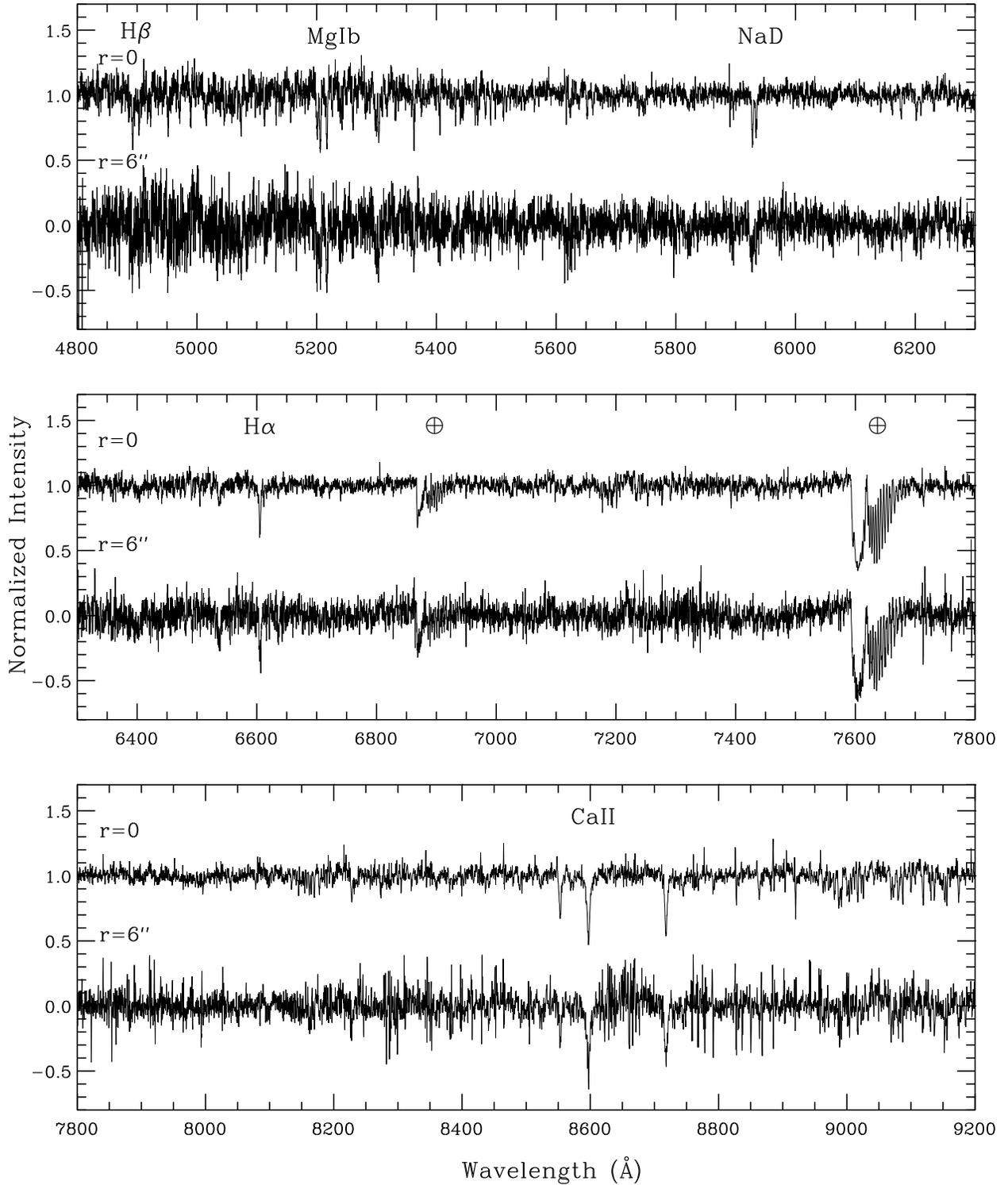}
\caption{Representative Keck/ESI spectra covering the continuous
wavelength region $\rm\lambda\lambda4800$--$9200\mbox{\AA}$ used in
determining kinematic profiles for our target galaxies.  Spectra of
VCC~1073 are shown for the central ($r=0''$) and outer ($r=6''$)
kinematic data positions.  The spectra are binned spatially to
$0.8''$, the size of the seeing disk at the time of observations, but
have not been smoothed in the spectral direction.  A few of the
important stellar absorption features in the dE are indicated, along
with the atmospheric A and B absorption bands.
\label{spec}}
\end{figure}

Data reduction was carried out on these rectified strips using
procedures similar to those routinely used on long-slit spectra.  Each
strip was divided by its corresponding normalized flat-field image
(from the same echelle order).  Cosmic rays on individual exposures,
identified on the basis of object sharpness and peak pixel brightness,
were masked and the exposures were then combined.  Each order was
logarithmically binned in wavelength using a two-dimensional
wavelength calibration (i.e.,~as a function of spatial position along
the slit) determined from a combined Cu/Ar/Hg/Ne arc lamp spectrum.
The rms of the residuals in the wavelength solution is $\rm
0.05\mbox{\AA}$ or less in each order.  The sky spectrum was
determined for each combined frame from a section near the end of the
slit farthest from the galaxy center ($r\sim15''$), and subtracted
from the rest of the two-dimensional spectrum.  We recognize that this
``sky'' spectrum is contaminated by light from the outer parts of the
target dE galaxy but this is unlikely to have a significant effect on
our results.\footnote{Based on the surface brightness profiles in
Figure~\ref{sbprof} we estimate that the ``sky'' spectrum is
contaminated by galaxy light at the level of $\lesssim1\%$ and
$\approx5\%$ relative to the dE's center, with and without the nuclear
contribution, respectively.  In the outer parts of the dE profile
($r=5''$), the contamination level is higher (20\%--30\%), but even
this should not be a problem since there are no strong radial
gradients in the mean velocity and velocity dispersion in the outer
parts of the galaxies.}  Bright, poorly-subtracted sky lines were
masked out during the kinematic fitting discussed below.  The galaxy
continuum flux in each order was then individually normalized to unity
using the IRAF CONTINUUM task.  A noise frame was created for each
galaxy spectrum which kept track of uncertainties in every pixel due
to read noise and Poisson noise taking into account the CCD gain and
number of readouts.  This noise frame is used as input into the
kinematic analysis of \S\,\ref{vp} as part of the formal error
calculations on the velocity profile parameters.  Finally, the strips
from the different echelle orders were combined, weighted by the noise
frame, to create a single two-dimensional long-slit spectrum.  Due to
low signal-to-noise in the reddest and bluest echelle orders, we do
not include them in the analysis; the final combined spectrum covers
$\rm\lambda\lambda4800-9200\mbox{\AA}$.  The spectra show no evidence
for gaseous emission lines at any radii.  Representative combined
galaxy spectra are shown in Figure~\ref{spec}.

The seeing FWHM during each spectroscopic observation was determined
by comparing the observed intensity profile along the ESI slit to high
spatial resolution $V$-band surface brightness profiles derived from
{\it HST\/} images (as discussed in \S\,\ref{sb}).  These brightness
profiles were convolved with a Gaussian seeing point spread function
(PSF), integrated over the $0.75''$ slit width and binned into
$0.18''$ pixels to match the ESI pixel scale in the spatial direction.
This was then compared to the observed intensity profile along the ESI
slit in the matching spectral region.  The best-fit Gaussian FWHM
seeing estimates are given in Table~2 for each galaxy.  These are
consistent with the less accurate estimates of the seeing FWHM
determined by coadding short (few second) exposures taken with the ESI
guider camera at intervals of approximately 5~minutes during the
spectroscopic observations.

\subsubsection{Measurement of Line-of-Sight Velocity and Velocity 
Dispersions}\label{vp}

The mean line-of-sight velocity and velocity dispersion as a function
of radius were determined using a pixel-fitting method first described
in \citet{vdm94}.  These quantities were determined by comparing the
observed galaxy spectrum to a stellar template convolved with a series
of Gaussian line profiles.  The best-fitting Gaussian profile was
determined by $\chi^2$ minimization in pixel space.  The free
parameters in this analysis are: mean line-of-sight velocity $v$,
velocity dispersion $\sigma$, and a line-strength parameter $\gamma$,
which measures the ratio of equivalent width in the galaxy to that of
the template star and accounts for template mismatch (e.g.,~due to
differences in effective temperature or metallicity).  Night sky
absorption features (A and B bands) and strong sky emission lines were
masked out in the fitting procedure; masked pixels were not included
in the calculation of $\chi^2$.  Deviations from Gaussian profiles
\citep{vdm93} were not fit as this requires higher signal-to-noise
spectra than presented in this paper.  In addition, an arbitrary
continuum term was simultaneously fit to the data approximated by the
sum of Legendre polynomials.  In analyzing the combined ESI spectra
($\approx 20,000$~pixels), the continuum was fit with a 20th order
polynomial.

We have tested this method on broadened template stars to determine
the minimum signal-to-noise required to accurately recover velocity
dispersions and to estimate our sensitivity to template mismatch.
Stellar templates were broadened with Gaussian kernels of varying
$\sigma$ and Poisson noise was added.  These tests suggest that for a
signal-to-noise level $\rm S/N \ge 10$ per spectral pixel, a galaxy's
internal velocity dispersions can be measured down to the
instrumental resolution of 23~km~s$^{-1}$~with an accuracy of 1\% and
down to 18.5\kms\ with an accuracy of 10\%.  We have spatially
rebinned our galaxy data to achieve a signal-to-noise level of $\rm
S/N \ge 10$ per pixel at all radii, while ensuring that the spatial
bin size is at least as large as the FWHM of seeing during the
observations ($\sim0.9''$).  Velocity profiles were recovered using
template stars ranging in spectral type G8III to M0III.  The best
fitting template, the K1III star HD 40460 ([Fe/H] = $-0.42$), was used
to recover all profiles presented here.  The choice of template star
did not affect the recovered profiles at a level in excess of the
formal error bars.  We also find no significant difference between the
profiles presented here to those determined by separately recovering
profiles for the eight ESI echelle orders and computing a profile
based on the weighted mean.

\subsubsection{A Reliability Test: Comparison of ESI and HIRES 
Data}\label{esi_hires_comp_sec}

\begin{figure}
\epsscale{0.8}
\plottwo{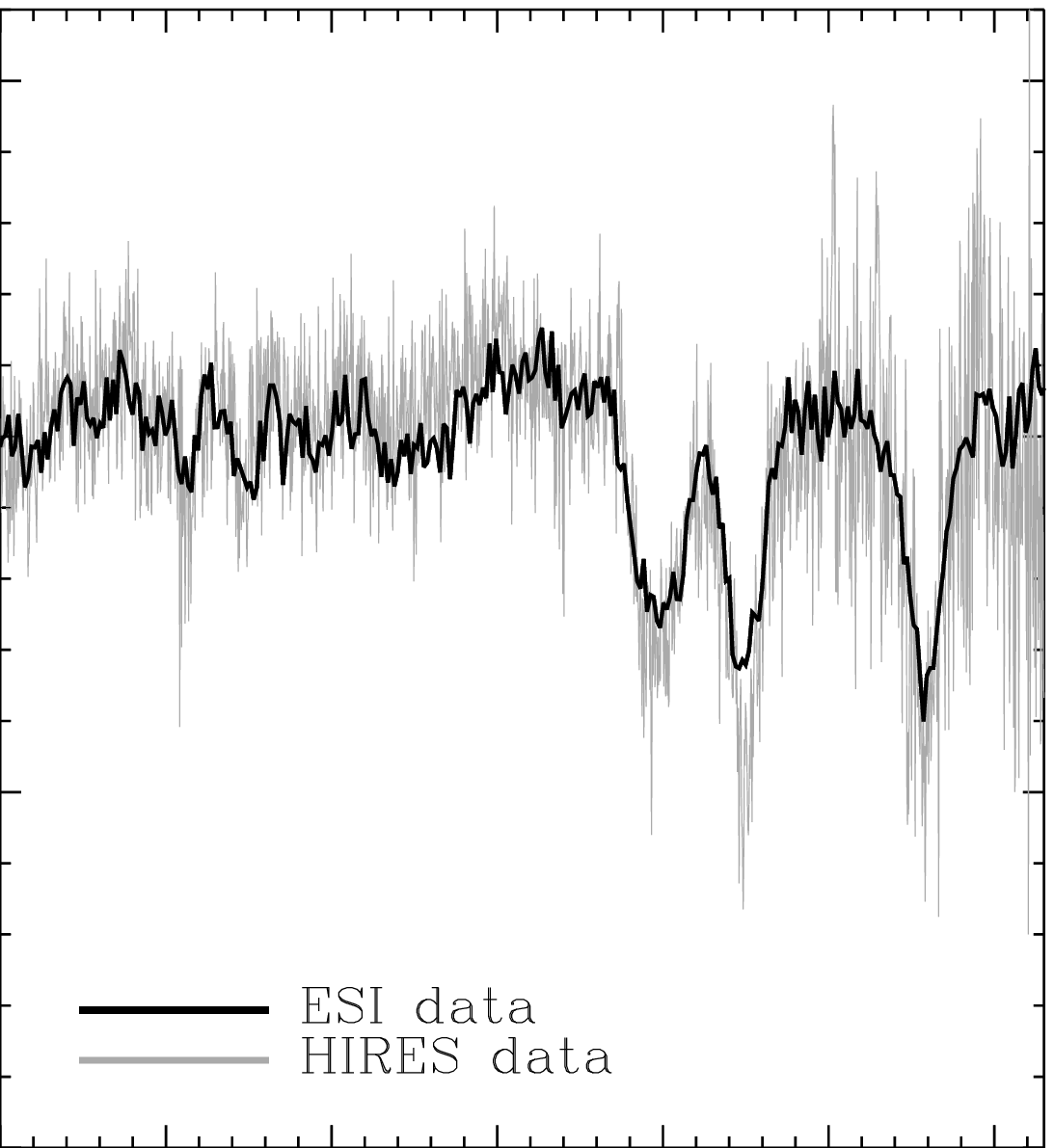}{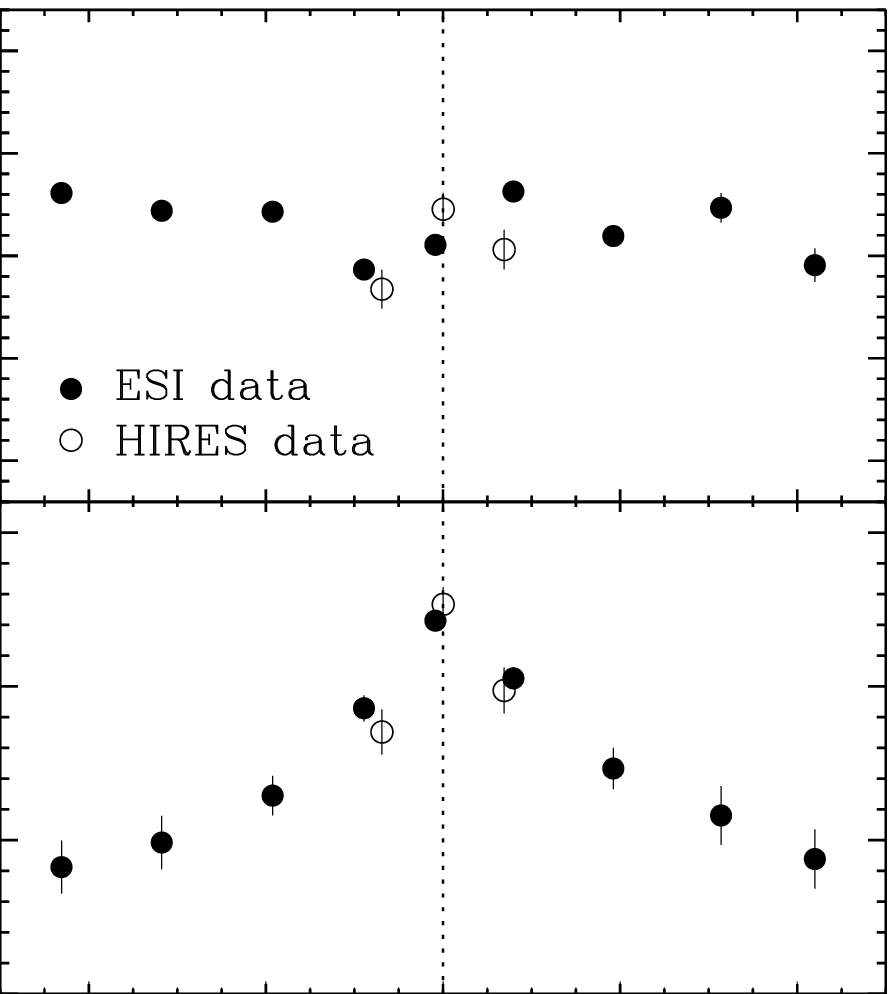}
\vskip 0.5 cm
\caption{{\it Left\/}: A spectral comparison of VCC~1254 in the region
of the MgIb lines as observed by the ESI (black line) and HIRES (grey
line) spectrographs on the Keck telescope.  The spectra cover the same
spatial region, the central $1''$ of the galaxy.  {\it Right\/}: Mean
radial velocity offset relative to the galaxy's systemic velocity
({\it upper right\/}) and line-of-sight velocity dispersion ({\it
lower right\/}) as a function of radius along the major axis as
determined by ESI (solid circles) and HIRES (open circles) along with
$1\sigma$ error bars, calculated in the wavelength region
$\rm\lambda\lambda5000$--$5250\mbox{\AA}$.  The two datasets are in
excellent agreement despite differences in instrumental resolution and
S/N ratio.\label{hires}}
\end{figure}

We have previously attempted to observe Virgo dE galaxies with
Keck/HIRES in March 1998.  The significantly higher spectral
resolution of HIRES (2.1~\kms~pixel$^{-1}$) and its lower throughput
as compared to ESI made these observations prohibitively difficult.
The dE galaxies VCC~1254, VCC~1073, VCC~452 and VCC~1876 were observed
for 5, 3, 2 and 1.8~hours, respectively, through a custom-made $2.0''
\times 11.0''$ slit.  We were able to determine the kinematics for
VCC~1254 inside $r<2''$; however, data for the last three~galaxies did
not have sufficient signal-to-noise to recover velocity profiles.
Although this HIRES spectrum does not provide additional information
on VCC~1254, it does provide an excellent reliability check on our ESI
observations.  The ESI and HIRES data complement each other in that
the ESI data have relatively high S/N but an instrumental resolution
approaching the intrinsic dispersion of our target galaxies, whereas
the instrumental resolution of HIRES is significantly higher at the
price of low signal-to-noise.  As shown in Figure~\ref{hires}, the
line profile shapes and velocity profiles determined with HIRES match
that measured by ESI.  The velocity profiles were calculated in the
wavelength region $\rm\lambda\lambda5000$--$5250\mbox{\AA}$, however,
we have no reason to believe that this argreement would be any less in
other spectral regions.

\subsection{Imaging}\label{img_sec}
\subsubsection{Observations and Data Reduction}\label{sb}

{\it HST\/} Wide Field Planetary Camera~2 (WFPC2) imaging is
available for each of our target galaxies.  These images provide high
spatial resolution surface brightness profiles needed for the
dynamical modeling discussed in \S\,\ref{models}, and allowed us
to measure photometric properties for our target galaxies.  The data,
first presented in \citet{mil98} and \citet{sti01}, consist of $2
\times 230$-s WFPC2 images in the F555W bandpass and a single 300-s
exposure in F814W.  The galaxies are centered on the WF3 CCD in the
{\it HST\/} pointings, and we use only the F555W WF3 CCD image to
determine surface brightness profiles.  The images were cleaned of
cosmic rays and combined.  The instrumental F555W magnitudes were
calibrated into $V$-band using the transformations of \citet{hol95},
assuming $(V-I)=1.0$ \citep{mil98}.  Surface brightness profiles were
determined for each galaxy using the IRAF ELLIPSE isophotal fitting
routine down to a surface brightness of $\mu_{V} \sim 24$.  The
average ellipticity, $\epsilon$, determined between $r=1''-20''$ and
the total integrated apparent magnitude, determined by integrating the
total flux inside a $40''$ aperture, are given in Table~1.  Our
apparent magnitudes agree with those determined by \citet{mil98}.
Unlike \citeauthor{mil98}, we consider VCC~1577 to be a nucleated dE
galaxy since its bright, central star cluster is within a few tenths of
an arcsecond of the galaxy's isophotal centroid position.  The observed
surface brightness profiles are shown in Figure~\ref{sbprof}.

\subsubsection{Surface Brightness Profile Fitting}\label{sbfits}

\begin{figure}
\plotone{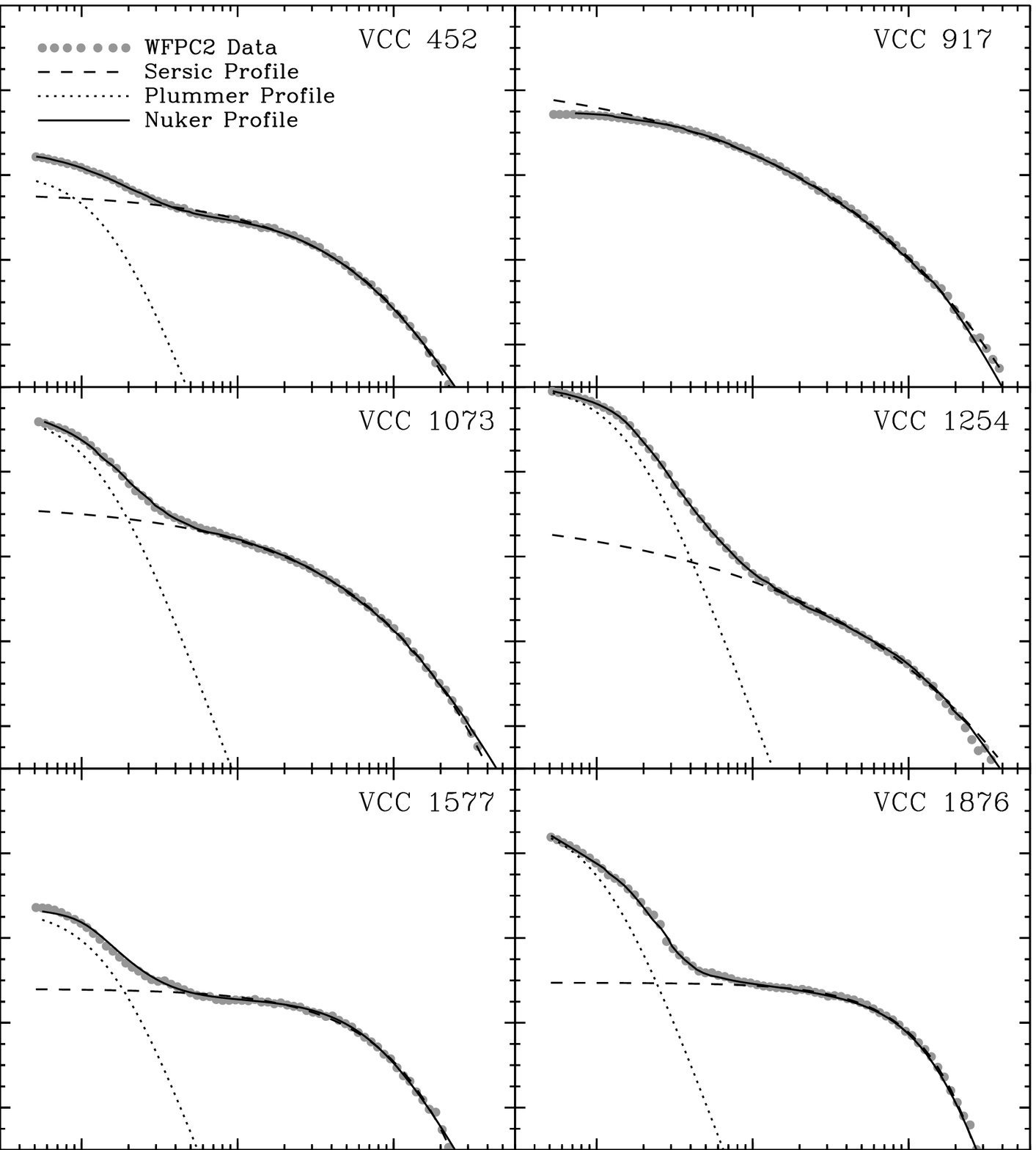}
\vskip 0.5 cm
\caption{Surface brightness profiles as a function of semimajor axis in the
$V$-band derived from {\it HST\/} F555W WFPC2 images for the six Virgo dE
galaxies (solid grey circles) and analytic profile fits: generalized
``Nuker'' law \citep{lau95} fit to the overall profile (solid line), Sersic
profile fit to the region outside $1''$ (dashed line), and Plummer profile
fit to the inner region of the nucleated galaxies taking into account the {\it HST\/}
WFPC2 PSF (dotted line; all except VCC~917).  Error bars on the
observed profiles are smaller than the data points.\label{sbprof}}
\end{figure}

In subsequent analysis, we differentiate between light from the
central dE nucleus and the underlying galaxy.  We therefore fit
separate analytic profiles to the inner and outer surface brightness
profiles.  For the underlying galaxy, a Sersic profile is fit of the
form: $I^{\rm gal}(r) = I_0^{\rm gal} {\rm exp}[(r/r_0)^{1/n}]$, where
a Sersic index $n=1$ represents an exponential profile and $n=4$ is a
de~Vaucouleurs law.  The best-fit Sersic profile is determined by
non-linear least-squares fitting to the region $r=1''-20''$; in this
region, contributions from the nucleus and effects of the {\it HST\/}
WFPC2 PSF should be negligible.  The resulting profiles are shown in
Figure~\ref{sbprof}, and Sersic indices and half-light effective radii
are listed in Table~3.  Best-fitting Sersic indices range between
$n=0.8 - 2.9$.  There is a slight trend towards smaller $n$ values
(more closely exponential) at fainter magnitudes, consistent with that
seen in the much larger sample of Virgo dE surface brightness profiles
analyzed by \citet{bin98}.

At the distance of the Virgo Cluster, the nuclei of dE galaxies are
slightly extended compared to the WFPC2 PSF.  We have used the ISHAPE
software developed by \citet{lar99} to derive shape parameters for
these nuclei.  The intrinsic shape of the nuclei was modeled as a
Plummer profile whose projected intensity scales as $I^{\rm nuc}(r) =
I_0^{\rm nuc}/[1 + (r/b_{\rm nuc})^2]^2$, where $b_{\rm nuc}$ is the
scale radius of the nucleus which, for this profile, is also the
half-light radius.  The ISHAPE software convolves the analytic profile
with the WPFC2 F555W PSF and a diffusion kernel generated by the
TinyTim software \citep{kri97}, and determines the best-fitting model
parameters by minimizing residuals between the model and original
two-dimensional image.  Nuclear profiles were fit inside the central
$1.0''$ (10 pixels) and were assumed to be circularly symmetric.  The free
model parameters are the effective radius, $b_{\rm nuc}$, the profile
normalization, $I_0^{\rm nuc}$, and a constant background level.
Although a constant background level is a good approximation for the
non-nuclear component inside the fitting radius, it is more
appropriate to subtract the underlying galaxy Sersic profile.
Therefore, we use the ISHAPE software to fit only the profile shape,
$b_{\rm nuc}$, and determine the profile normalization, $I_0^{\rm
nuc}$, such that the total nuclear magnitude equals the luminosity
leftover after subtraction of the galaxy Sersic profile from the total
observed surface brightness profile.  For the five nucleated dEs, the
resulting nuclear profiles are plotted as dotted lines in
Figure~\ref{sbprof}; effective half-light radii and total nuclear magnitudes
are given in Table~3.

\section{Results}\label{res_sec}

\begin{figure}
\plottwo{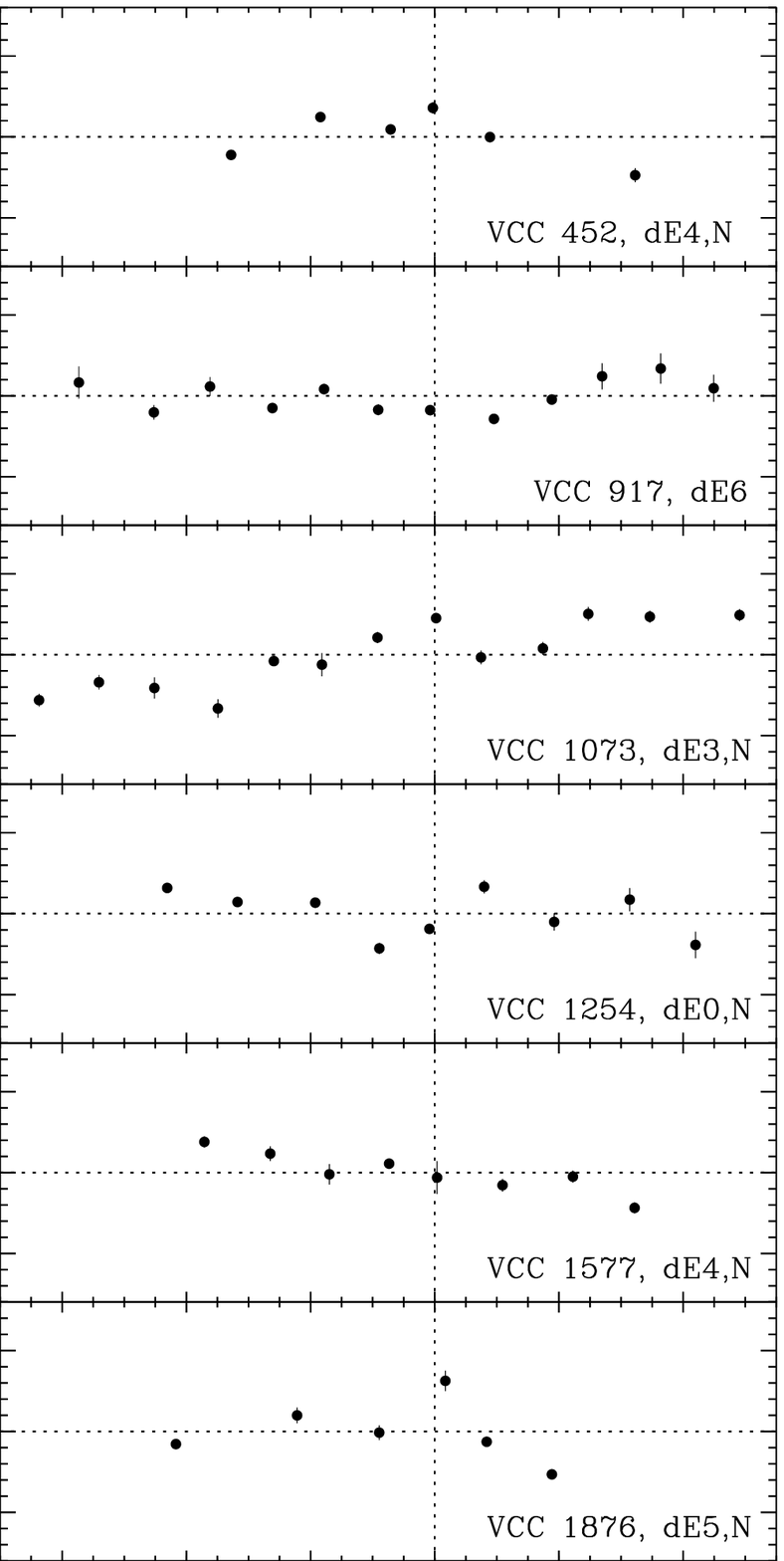}{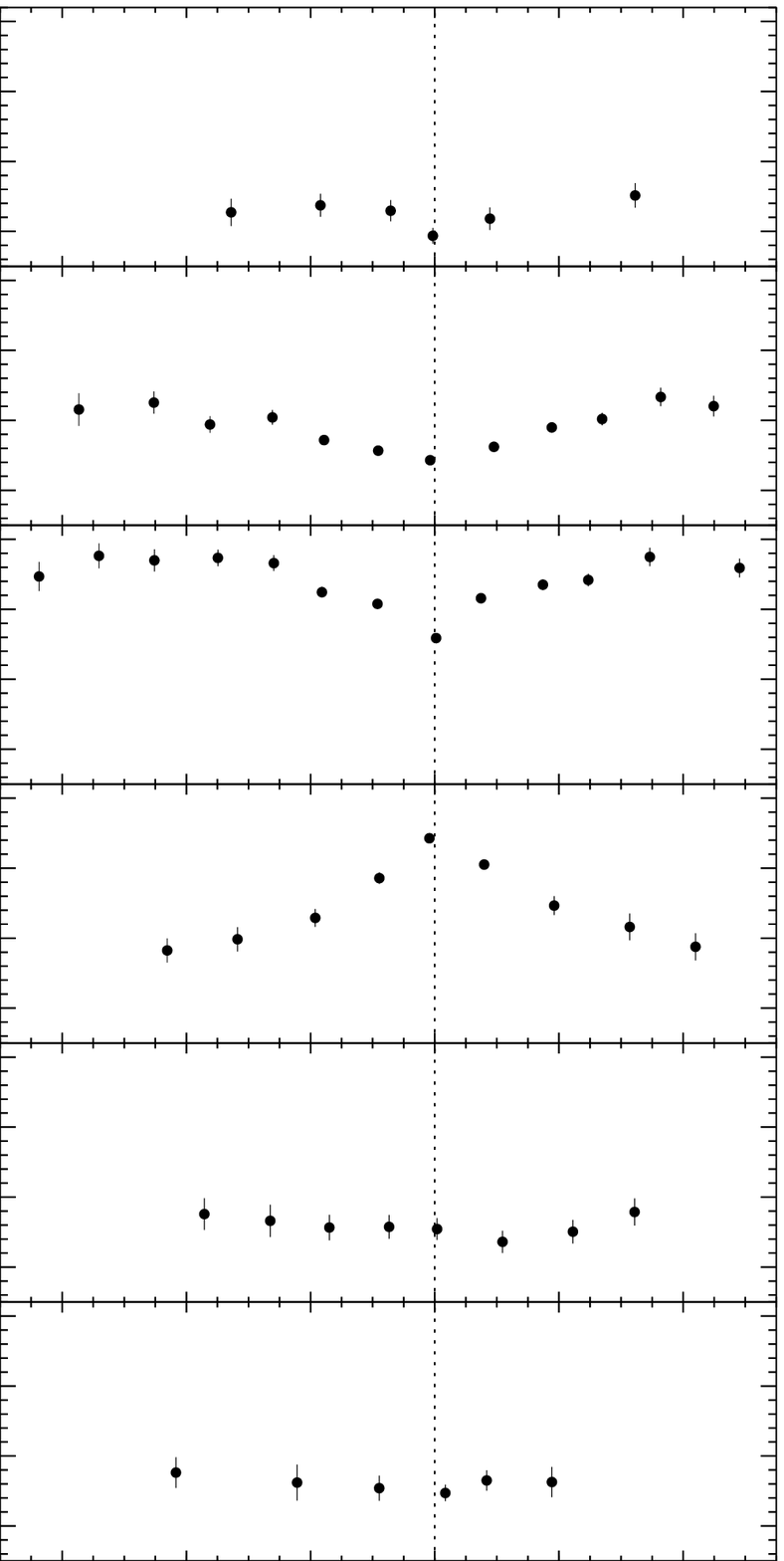}
\vskip 0.5 cm
\caption{The mean line-of-sight velocity offset relative to the galaxy's
systemic velocity ({\it left panel\/}) and velocity dispersion ({\it right
panel\/}) as a function of radial distance along the major axis for the six
Virgo dEs observed with ESI.  At the distance of Virgo, $1''$ corresponds to
$\sim100$~pc.  None of the dEs shows significant rotation with the possible
exception of VCC~1073 and VCC~1577.  The nuclear velocity dispersion is
greater than that of the underlying galaxy in the case of VCC~1254, and
smaller in the cases of VCC~452 and VCC~1073.\label{vp_fig}}
\end{figure}

Results of the kinematic analysis of \S\,\ref{vp} are shown for all
galaxies in Figure~\ref{vp_fig}; the derived kinematic profiles are
summarized in Table~4.  The mean line-of-sight velocity and velocity
dispersion are plotted as a function of major axis radius in
arcseconds; the radius was measured relative to the peak position of
the intensity profile along the ESI slit.  The systemic radial
velocity of each dE was determined from the mean of the velocity
data points and subtracted from the velocity profile.  The corrected
heliocentric velocities are listed in Table~4 and agree, within
measurement errors, with previously published radial velocity
measurements for VCC~917, VCC~1073, VCC~1254 and VCC~1876
\citep*{bin85}.  

\subsection{Velocity Profiles: A Lack of Rotation}\label{rot}

\begin{figure}
\plotone{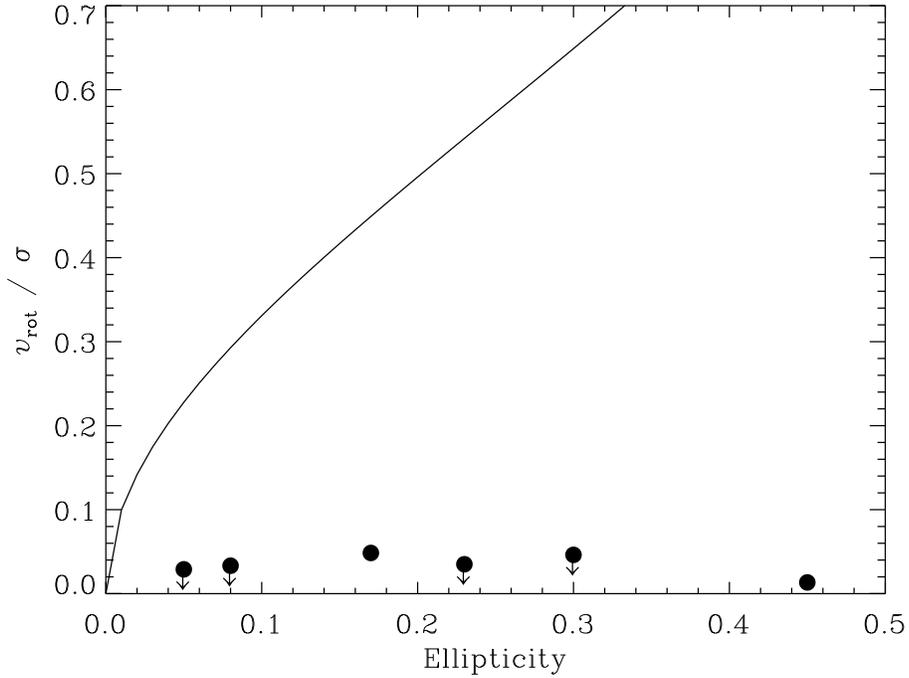}
\vskip 1 cm
\caption{The ratio of the upper limit on the rotation velocity $v_{\rm
rot}$ to the observed velocity dispersion $\sigma$ plotted versus mean
isophotal ellipticity for six Virgo dE galaxies.  The symbols without
upper limit arrows are the two dEs which appear to have
some rotation, VCC~1073 and VCC~1577. The solid line is the expected
relation for an oblate, isotropic galaxy flattened by rotation, seen
edge-on; systems that are not edge-on should have somewhat larger
predicted $v_{\rm rot}/\sigma$ values (\S\,\ref{rot}).  The six dEs in
our sample do not appear to be rotationally
flattened.\label{rotation}}
\end{figure}

The velocity profiles in Figure~\ref{vp_fig} show no evidence for
substantial rotation along the major axis of any of the six dE systems
observed.  To quantify the maximum rotation velocity allowed by the
data, we have differenced the average velocities on either side of the
major axis of the galaxy and divided by two ($v_{\rm rot}$ in
Table~4).  This quantity is not particularly meaningful for galaxies
which do not show a coherent rotation curve; it merely represents an
upper limit on rotational motion.  Error bars on rotational motion
were determined by adding in quadrature the error of the mean velocity
on either side of the major axis.  If the observed flattening of these
galaxies were determined by rotational motion alone, the expected
rotation velocity can be calculated directly from the tensor virial
theorem, given the observed velocity dispersion \citep{bin87}.  The
ratio of the maximum rotational velocity to the average velocity
dispersion ($v_{\rm rot}/\sigma$) is plotted versus ellipticity in
Figure~\ref{rotation} and compared to the ratio expected from an
isotropic, rotationally-flattened body.  Ellipticity and average
velocity dispersion are determined outside $r=1''$ in order to exclude
any contributions from a central nucleus and are listed in Tables~2
and 3, respectively.  The upper limits on $v_{\rm rot}/\sigma$
determined for these galaxies are significantly smaller than expected
if the observed flattenings were due to rotation.  Thus, from
Figure~\ref{rotation} we conclude that these dEs are primarily
flattened by anisotropic velocity dispersions.

The low rotation velocities of this study are consistent with previous
measurements of dE velocity profiles in the Local Group \citep{ben90}.
However, recent studies of dE kinematics suggest that a fraction of dE
galaxies are rotationally supported.  A rotational velocity of 15
\kms~was measured by \citet{der01} for FS~76, a dE in the Fornax
Cluster, slightly below the value expected if this system is
rotationally supported.  \citet{ped02} found that five of their six
Virgo Cluster dEs rotated with maximum velocities between 15 to 30
\kms, placing them on or above the relation for rotational support.
The sixth galaxy presented by \citeauthor{ped02}, IC~794 or VCC~1073,
is also in our sample.  Their measurement, $v_{\rm rot} = 3.4 \pm 1.7$
\kms, is consistent with our observation of low rotation velocity,
$v_{\rm rot} = 2.1 \pm 0.4$\kms.  Thus, of the 11 Virgo dEs with
measured velocity profiles, 5 have significant rotation velocities and
6 are non-rotating.  We note that the average brightness of the
rotating dE sample ($M_V = -17.6$, assuming (B-V) = 0.8 \citep{gav01})
is slightly brighter than that of our non-rotating dEs ($M_V =
-16.4$).  In \S\,\ref{fp} we find that these two populations are also
slightly separated in the Fundamental Plane.  We have recently
observed a larger sample of Virgo dE galaxies, some of which have
significant rotation velocities and some which do not.  We will
explore in more depth the differences between these two classes in a
forthcoming paper.

\subsection{Interpreting Velocity Dispersion Profiles}\label{interp_vp_sec}

The mean velocity dispersions of the six dE galaxies presented in
Figure~\ref{vp_fig} lie between 20 and 55~\kms, and show a wide range
of profile shapes.  Although Virgo dE nuclei are unresolved from the
ground, the kinematics of the central nucleus appear to be
distinguished from the underlying galaxy.  Surprisingly, the central
velocity dispersion can be either larger (VCC~1254) or smaller
(VCC~1073, VCC~452) than the surrounding galaxy.  Below, we construct
dynamical models for each of the observed galaxies in order to explore
the range of mass-to-light ratios, velocity dispersion anisotropy and
central black hole masses allowed by the observed profiles.

\subsubsection{Dynamical Modeling}\label{models}

High spatial resolution WFPC2 imaging available for all of the
observed dEs (\S\,\ref{sb}) allows dynamical modeling of the velocity
dispersion profiles through the assumption that the stellar mass
density is proportional to the luminosity density times some
mass-to-light ratio at all radii.  Solving the spherically symmetric
Jeans equation, the predicted kinematics are convolved through the
observational setup, allowing a direct comparison to the observations.
The simplifying assumption of spherical symmetry is justified as more
generalized models cannot be discriminated against without additional
information such as minor axis kinematics or higher order velocity
profile moments.  We produce spherical models in which the radius is
related to the observed semimajor/semiminor axes and ellipticity
of the corresponding dE galaxy via the relation:
$r=\sqrt{ab} = a \sqrt{1-\epsilon}$.  The square root of the product
of the semi-major and minor axis is a more appropriate quantity than
the semi-major axis for these elliptical systems.  The models are
based on dynamical software described in more detail by \citet{vdm94}.

The luminosity density, $j(r)$, is determined for each galaxy by Abel
transformation of the projected WFPC2 $V$-band surface brightness
profile measured in \S\,\ref{sb}.  To avoid noise amplification, the
observed surface brightness profiles were first fit to an arbitrary
function, a generalization of the ``Nuker law'' \citep{lau95}, as
shown in Figure~\ref{sbprof}.  The total luminosity density is assumed
to be composed of two parts, $j(r) = j_{\rm nuc}(r) + j_{\rm gal}(r)$,
representing the central nucleus and underlying galaxy, respectively.
We model the nuclear component as a Plummer model with fixed scale
length, $b_{\rm nuc}$, and total nuclear luminosity, $L_{\rm nuc}$ in
the $V$ band, as determined for each galaxy from the projected
luminosity density in \S\,\ref{sbfits}.  The luminosity density of a
Plummer model, taken from \citet{dej87}, is:
\begin{equation}\label{jn}
j_{\rm nuc}(r) = \frac{3 L_{\rm nuc}}{4 \pi b_{\rm nuc}^3}~\Big{[}1 + 
\frac{r^2}{b_{\rm nuc}^2}\Big{]}^{-5/2}
\end{equation}
\noindent
The luminosity density of the underlying galaxy, $j_{\rm gal}(r)$, is
the total luminosity density minus the contribution from the nucleus.
The mass density of stars, $\rho(r)$, is modeled as the luminosity
density times a mass-to-light ratio.  In our models, we allow two
distinct mass-to-light ratios as free parameters, $\Upsilon_{\rm gal}$
and $\Upsilon_{\rm nuc}$, for the underlying galaxy and nuclear
component.  The mass density distribution of the galaxy component is
then: $\rho_{\rm gal}(r) = \Upsilon_{\rm gal} \> j_{\rm gal}(r)$.  The
mass density distribution of the nucleus is similarly defined as:
$\rho_{\rm nuc}(r) = \Upsilon_{\rm nuc} \> j_{\rm nuc}(r)$.  The total
mass density of the system is the sum of these densities: $\rho(r) =
\Upsilon_{\rm nuc} \> j_{\rm nuc}(r) + \Upsilon_{\rm gal} \> j_{\rm
gal}(r)$.  Rearranging terms, the total density is modeled as:
\begin{equation}\label{rho} 
\rho(r) = \Upsilon_{\rm gal} \> j(r) + [(\Upsilon_{\rm nuc} -
\Upsilon_{\rm gal}) \> j_{\rm nuc}(r)]
\end{equation}
\noindent
where the total luminosity density $j(r)$ is inferred from the {\it
HST\/} WFPC2 surface brightness profile, $j_{\rm nuc}(r)$ is given in
Eqn.~(\ref{jn}), and the mass-to-light ratios $\Upsilon_{\rm nuc}$ and
$\Upsilon_{\rm gal}$ are free parameters.  The case where
$\Upsilon_{\rm nuc} = \Upsilon_{\rm gal}$ is equivalent to the mass
density of the galaxy being equal to the luminosity density times a
constant mass-to-light ratio.

The total gravitational potential, $\Phi$, is obtained by integrating
over the mass density determined by Eqn.~(\ref{rho}) plus an added
possible contribution, $GM_{\rm BH}/r$, from a central black hole.
The velocity dispersion as a function of radius, $\sigma_r$, is then
obtained by solving the spherically-symmetric Jeans equation:
\begin{equation}
\frac{d(\rho \sigma_{r}^2)}{dr} + 2 \frac{\beta \rho \sigma_{r}^2}{r}
= -\rho \frac{d\Phi}{dr}
\end{equation}
\noindent
where $\beta = 1 - \sigma_{\theta}^2 / \sigma_{r}^2$ describes the
velocity dispersion anisotropy.  Models with $\beta =0$ are isotropic,
$\beta < 0$ are tangentially anisotropic, and $0 < \beta \le 1$ are
radially anisotropic.  The radial velocity dispersion $\sigma_r$ is
numerically evaluated for any combination of $\beta$, $M_{\rm BH}$,
$\Upsilon_{\rm gal}$, $\Upsilon_{\rm nuc}$, and surface brightness
profile.  In order to compare to observations, the velocity
dispersions are projected along the line-of-sight.  The projected
dispersions are convolved using Monte Carlo integration with a
Gaussian kernel to take into account the seeing FWHM (as determined in
\S\,\ref{datared_sec}).  The dispersions are then sampled using the
slit width, pixel size, and rebinning scheme specific to each
observation.  The predicted dispersion $\sigma(r)$ can be compared
directly to the observed velocity dispersions $\sigma_i$ over all
radial bins $r_i$ by the defined quantity:
\begin{equation}\label{chi}
\chi_{\sigma}^2=\Big{[}\frac{\sigma_i-\sigma(r_i)}{\Delta\sigma_i}\Big{]}^2
\end{equation}
\noindent
The best-fitting model is determined by minimization of
$\chi_{\sigma}^2$.  Given the well known degeneracy between the mass
profile and velocity anisotropy, and given the quality of our data, we
choose not to explore the full range of allowed parameter space.
Instead, we consider three limiting cases: (1)~constant mass-to-light
ratio and velocity anisotropy, (2)~the same model plus a central black
hole, and (3)~models with separate mass-to-light ratios for the
nucleus and underlying galaxy and no central black hole.

\subsubsection{Mass-to-Light Ratios and Orbital
Anisotropy}\label{ml_aniso_sec}

\begin{figure}
\plotone{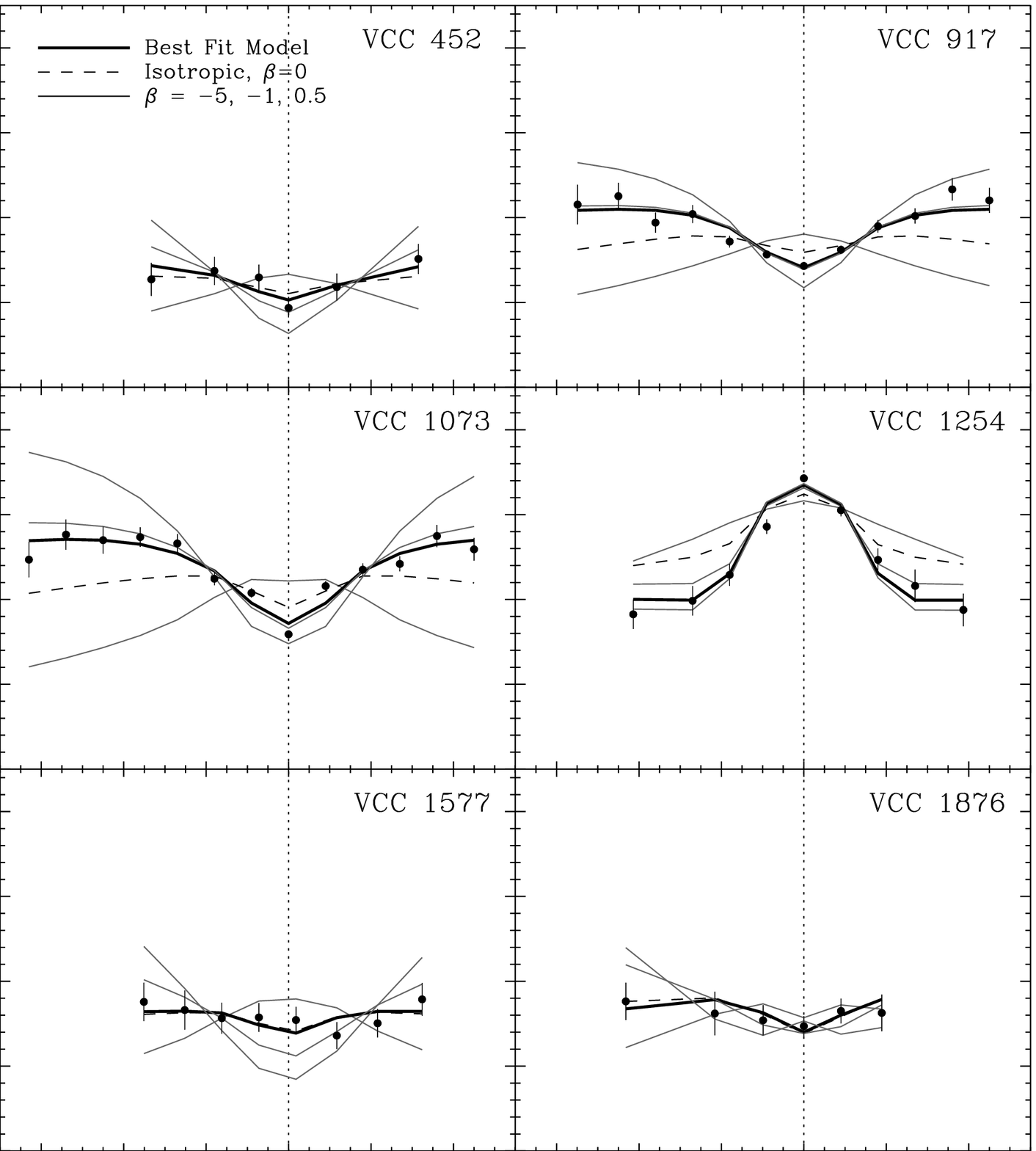}
\vskip 0.5 cm
\caption{Observed line-of-sight velocity dispersion as a function of
major axis radius (points with $\pm1\sigma$ error bars), compared to
the predictions of dynamical models with constant velocity anisotropy.
The thick solid line indicates the best fit model.  For each galaxy,
models with $\beta=-5$, $-1$, and 0.5 (thin solid lines) and isotropic
$\beta=0$ models (dashed lines) are indicated.  The global
mass-to-light ratio is optimized via minimization of $\chi_{\sigma}^2$
for each $\beta$ value (\S\,\ref{ml_aniso_sec}).  The first four
galaxies are fit by tangentially-anisotropic models, while the last
two are fit by nearly isotropic models.\label{aniso_models}}
\end{figure}

We first consider models without a central black hole for which the
free parameters are a single mass-to-light ratio $\Upsilon_V$
(i.e.~$\Upsilon_{\rm gal} = \Upsilon_{\rm nuc}$) and velocity
dispersion anisotropy, $\beta$, both independent of radius.  For each
galaxy, models were run for values of the velocity dispersion
anisotropy ranging between $-3 \le \beta \le 0.75$.  Best fitting
values of $\beta$ and $\Upsilon_V$ were determined by overall
minimization of $\chi_{\sigma}^2$ and are listed in Table~4.  Formal
$1\sigma$ (68\% confidence) error bars are calculated for each
individual free model parameter by the variation needed to increase
$\chi_{\sigma}^2$ by 1 with respect to its minimum value
\citep{pre92}.  Best-fit models, as well as several representative
$\beta$-value models are plotted over the observed data points in
Figure~\ref{aniso_models}.  Although isotropic models ($\beta=0$) do
not fit the observed profiles in detail, such models do in general
reproduce the dip or rise in the central dispersion observed in all
six~dE galaxies.  Most of the galaxies are best fit with tangential
anisotropy.  In some cases, the amount of anisotropy needed to fit the
profiles is unphysically large, motivating the models described in
\S\,\ref{nuc_vs_gal_sec}.

\begin{figure}
\plotone{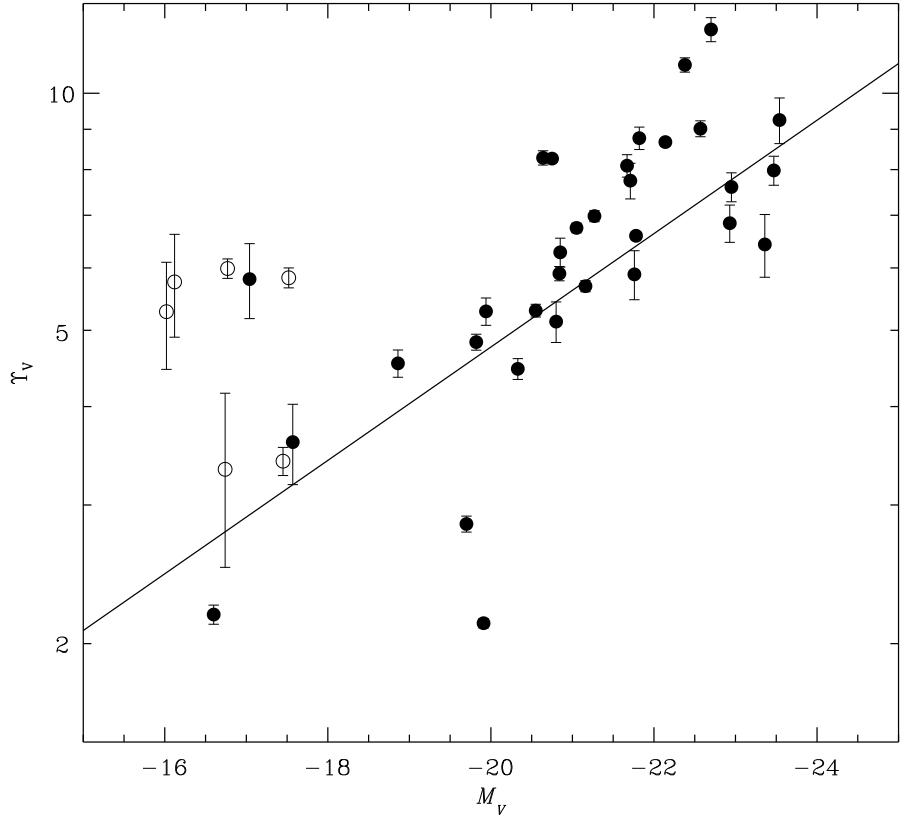}
\vskip 1 cm
\caption{Mass-to-light ratio in the $V$-band $\Upsilon_V$ versus total
absolute magnitude for the observed Virgo dE galaxies (open circles)
as compared to higher luminosity classical ellipticals (solid circles)
of \citet{mag98}.  The solid line indicates the Fundamental Plane correlation
$\Upsilon_V\propto L_V^{0.2}$ fit to the classical
ellipticals.\label{fig_ml}}
\end{figure}

The $V$-band mass-to-light ratios determined for our six Virgo dEs
range between $3~\le~\Upsilon_V~\le~6$.  These mass-to-light ratios
are plotted against the absolute total magnitude of each galaxy in
Figure~\ref{fig_ml} and compared to higher luminosity classical
elliptical galaxies of \citet{mag98}.  The \citeauthor{mag98} galaxies
show a clear trend of decreasing mass-to-light ratio towards fainter
magnitudes.  The observed dEs tend to have larger mass-to-light ratios
at a given absolute magnitude than expected by extrapolation of this
relationship.  

Combining the WFPC2 colors of these dE galaxies with the dynamically
determined mass-to-light ratios, it is possible to roughly determine
ages and metallicities for these galaxies.  The colors of the six dEs
lie in the range $1.0 \le (V-I) \le 1.2$, as measured by \citet{sti01}
from WFPC2 data averaged inside $r \le 10''$.  Approximating the
stellar populations of these dEs galaxies by a single-burst population
\citep{wor94}, the ages and metallicities implied by the above
combined constraints lie between 5 to 12~Gyr and $-1 \le \rm [Fe/H] \le
0$~dex, respectively.  This rough calculation suggests that the mass
determined from the observed kinematics can be accounted for by
stellar populations alone without the need for a significant dark
matter component, at least inside the radius of our observations
($\approx 1$ effective radius).  This does not rule out a significant
dark matter component at larger radii.  Accurate determination of the ages
and metallicities of these galaxies requires a rigorous analysis of
their line strengths and will be presented in a forthcoming paper.

\subsubsection{Upper Limits on the Mass of a Central Black
Hole}\label{bh_sec}

\begin{figure}
\plotone{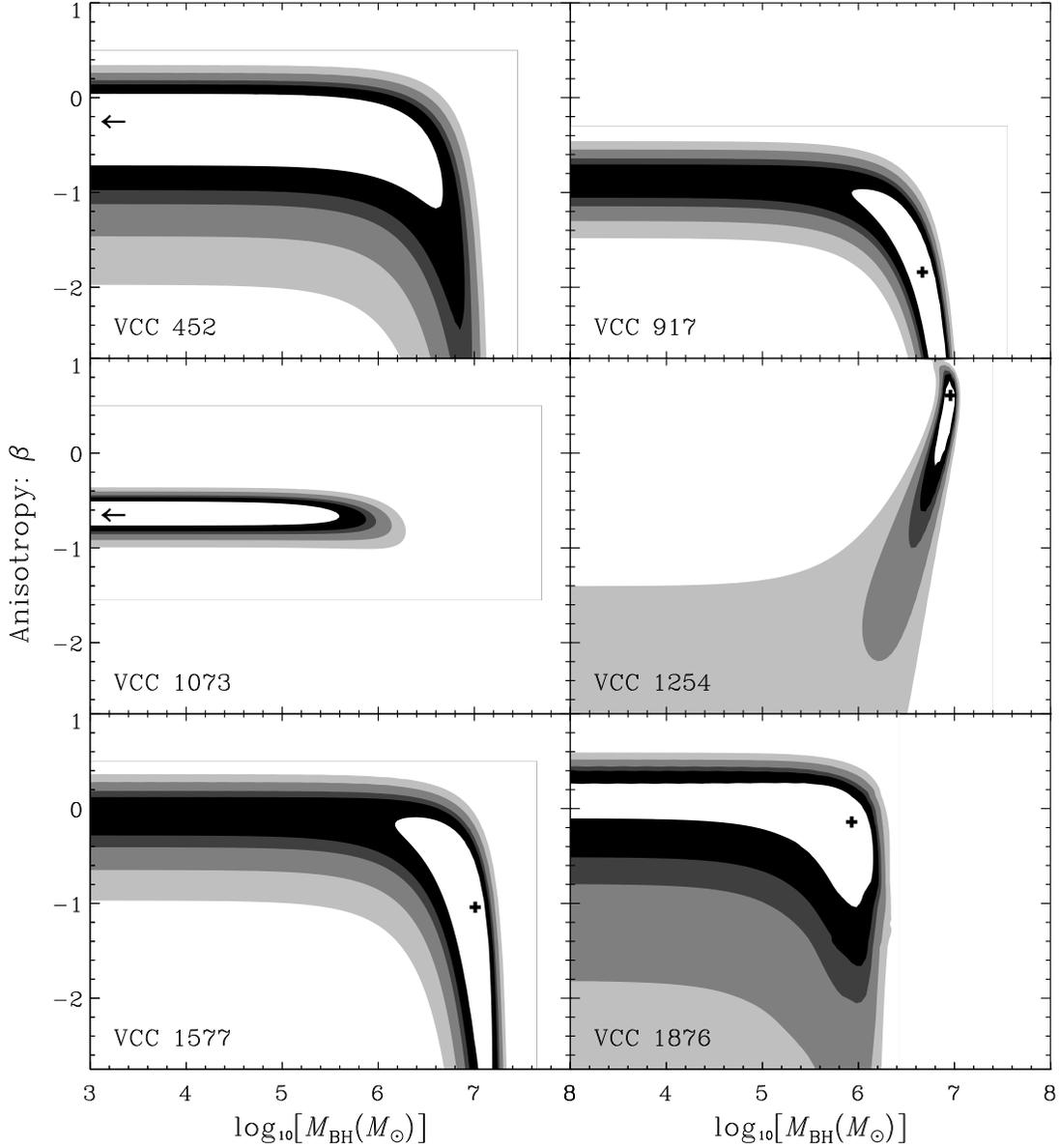}
\vskip 1 cm
\caption{Confidence contours for our sample of six Virgo dE galaxies
based on fits of dynamical models with constant velocity anisotropy
$\beta$ and a central black hole mass (\S\,\ref{bh_sec}): 68\%, 90\%,
95\%, 99\%, and 99.9\%.  The ``+'' symbol indicates the best fitting
model parameters; an arrow indicates that the best fit is for a zero
black hole mass.  Black hole masses greater than $M_{\rm
BH}>10^{7}\Msun$ can be ruled out at the 99.9\% confidence level for
all galaxies.\label{bh_models}}
\end{figure}

\begin{figure}
\plotone{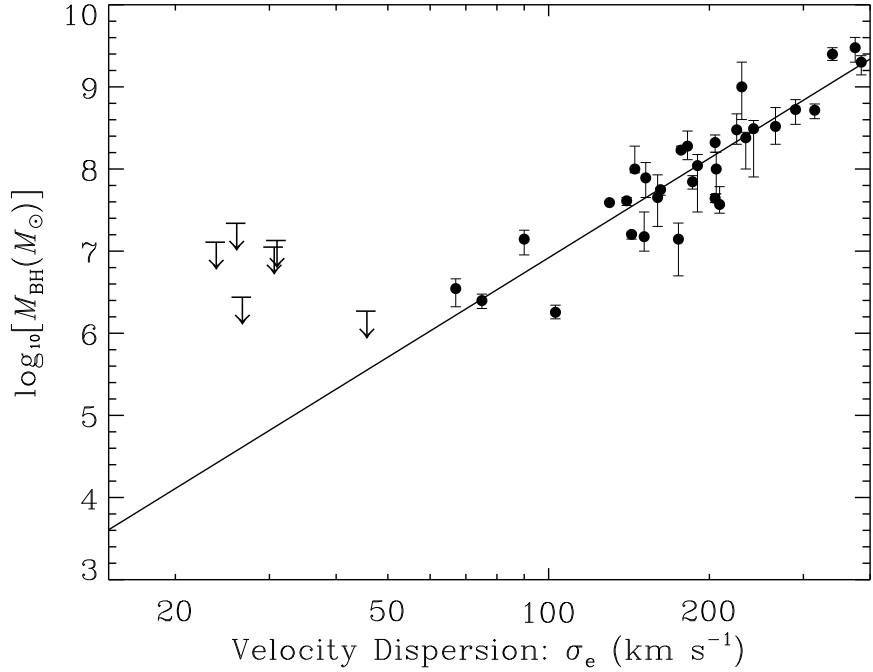}
\vskip 1 cm
\caption{Upper limits on the central black hole mass for the observed Virgo dE
galaxies (downward arrows) compared to the relationship between black hole
mass and bulge velocity dispersion $\sigma_e$ (solid line) determined by
\citet{tre02} for ellipticals and bulge-dominated spiral galaxies (solid
circles).  The upper limits on $M_{\rm BH}$ for our Virgo dE galaxies are
consistent with the extrapolation of the linear relation fit to the more
luminous early-type systems.\label{fig_sigBH}}
\end{figure}

We next allow an additional free model parameter in the form of a
central black hole with the goal of placing upper limits on the black
hole mass allowed by our kinematic data.  If, for example, dEs are the
morphologically-transformed remnants of larger progenitor galaxies,
limits on the central black hole mass can place potentially
interesting constraints on such a progenitor population.  Models were
run for a two-dimensional grid of velocity anisotropy versus black
hole mass.  For each grid point, the best fitting mass-to-light ratio
is determined.  Contours of constant $\Delta \chi_{\sigma}^2$ are
shown for these two parameters in Figure~\ref{bh_models}.  Confidence
intervals were assigned to $\Delta \chi_{\sigma}^2$ values in a two
dimensional parameter space, as discussed in \citet{pre92}.  Black
hole masses greater than $M_{\rm BH} > 10^{7} \Msun$ can be ruled out
at the 99.9\% confidence level.  For most objects, a zero mass black
hole model is either the best fitting model, or statistically similar
to the best fit at the 90\% confidence level ($1.7\sigma$).  The
galaxy VCC~1254 is the only dE in which a non-zero black hole mass,
$M_{\rm BH} = 9\times10^{6} \Msun$, is a significantly better fit to
the data than models without a central black hole.  This does not
necessarily imply the presence of a black hole, as we will show below
that this profile is equally well fit by a model in which the
mass-to-light ratio of the nucleus is larger than the underlying
galaxy.  In addition, although the upper limits on black hole mass
determined in this section are robust, actual black hole mass
determinations would require more complicated models which allow
$\beta$ variations with radius.  Upper limits on central dE black hole
mass are compared, in Figure~\ref{fig_sigBH}, to the black hole
mass-bulge velocity dispersion relationship, $M_{\rm BH}-\sigma_e$,
derived for bulge-dominated galaxies \citep*{geb00,fer00,tre02}.
Although this relationship may not be applicable to dE galaxies, which
lack a bulge component, our upper limits are still consistent with the
relationship.  The implication of these upper limits on possible dE
progenitor galaxies is discussed in \S\,\ref{disc_sec}.

\subsubsection{Nuclear versus Galaxy Mass-to-Light
Ratios}\label{nuc_vs_gal_sec}

The kinematic profiles of the five nucleated dEs in Figure~\ref{vp_fig}
show that the nuclei tend to have velocity dispersions
distinct from the surrounding galaxy.  In addition, photometric
studies suggest that dE nuclei tend to have different colors than the
underlying light of the host dE galaxy \citep*{sti01,dur97}.
Motivated by these observations, we consider models allowing two
distinct mass-to-light ratios, one for the nucleus, $\Upsilon_{\rm
nuc}$, and another for the underlying galaxy, $\Upsilon_{\rm gal}$.
We explore only isotropic models ($\beta = 0$) and search for
combinations of $\Upsilon_{\rm gal}$, $\Upsilon_{\rm nuc}$ which
minimize $\chi_{\sigma}^2$.  For VCC~1254, in which the nuclear
dispersion is larger than the surrounding galaxy, the best fit nuclear
mass-to-light ratio is twice that for the surrounding galaxy
($\Upsilon_{\rm nuc} = 2.1 \Upsilon_{\rm gal}$).  This model fits the
data equally well as the models presented in the previous two
sections.  However, a larger nuclear mass-to-light ratio implies an
older, and therefore redder, stellar population, contrary to
observations that the nucleus of VCC~1254 is bluer than the
surrounding galaxy \citep*{sti01,dur97}.  The true dynamical state of
VCC~1254 is likely to lie between the three extreme models presented
in this and previous sections.

For the four nucleated galaxies in which the velocity dispersion dips
in the central regions (VCC~452, VCC~1073, VCC~1577, and VCC~1876), a
smaller nuclear mass-to-light ratio ($\Upsilon_{\rm nuc} <
\Upsilon_{\rm gal}$) is only a marginally better fit to the data.  In
these four systems, we are not able to directly constrain the nuclear
mass-to-light ratio.  Even the unphysical case in which the nucleus
contributes no mass, isotropic models are inadequate fits to the
profiles of VCC~452 and VCC~1073, implying that these galaxies must
have some degree of tangential velocity anisotropy.  Isotropic, single
mass-to-light ratio models are adequate fits to the profiles of
VCC~1577 and VCC~1876, and variations in $\Upsilon_{\rm nuc}$ do not
significantly improve the fit.  This can be understood because the
nuclear component does not dominate the observed spectroscopic light
of these galaxies, not even in the central data point.  Inside $r<1''$,
the nucleus contributes between 5\% and 25\% of the total light
in this region, as compared to 60\% for VCC~1254.  Thus, the
measured central velocity dispersion is not a good estimate of the
velocity dispersion of the nucleus.  In order to place these nuclei on
the Fundamental Plane (see \S\,\ref{fp_nuc}), we do need an estimate
of the nuclear velocity dispersions.  For this, we assume a nuclear
mass-to-light ratio equal to the galaxy ($\Upsilon_{\rm nuc} =
\Upsilon_{\rm gal}$) and calculate the central projected velocity
dispersion for a Plummer model \citep{dej87} with total luminosity and
scale radius as determined in \S\,\ref{sbfits}.  The resulting nuclear
velocity dispersions are given in Table~4.

\subsection{The Fundamental Plane}\label{fp}

In the multivariate space defined by central velocity dispersion,
$\sigma_0$, effective surface brightness, $\mu_{\rm eff}$, and
effective radius, $r_{\rm eff}$, dE galaxies occupy a region of the
so-called Fundamental Plane distinct from classical elliptical
galaxies.  The separation is best demonstrated by the $\kappa$-space
projection of this parameter space defined by \citet*{ben92} as:
\begin{equation}
\kappa_1 \equiv (\log [\sigma_0^{2}] + \log r_{\rm eff}) / \sqrt2
\end{equation}
\begin{equation}
\kappa_2 \equiv (\log [\sigma_0^{2}] + 2 \log I_{\rm eff} - \log r_{\rm
eff})/\sqrt6
\end{equation}
\begin{equation}
\kappa_3 \equiv (\log [\sigma_0^{2}] - \log I_{\rm eff} - \log r_{\rm
eff})/\sqrt3
\end{equation}
\noindent
where $I_{\rm eff}$ is defined as $10^{-0.4(\mu_{\rm eff}-27)}$ and
is the mean intensity inside the radius $r_{\rm eff}$.  These
coordinates are related to physical quantities as follows: $\kappa_1$
is proportional to the logarithm of mass, $\kappa_2$ is proportional
to the effective surface brightness times mass-to-light ratio and
$\kappa_3$ is proportional to the logarithm of mass-to-light ratio.
To compare the location of our dEs in the Fundamental Plane to other
galaxy types, we plot data compiled by \citet*{bur97} for classical
ellipticals, spiral bulges, previously observed dEs, dwarf spheroidals
and globular clusters.  These data have been compiled in the $B$-band.
For comparison, we transform our dE data to the $B$-band
assuming $(B-V) = 0.8$ \citep{gav01}.  In addition, we add four of the
five rotating dE galaxies presented by \citet{ped02} for which
photometric data is available.  Photometric properties for
UGC~7436/VCC~543 were determined from WFPC2 imaging as in
\S\,\ref{sbfits}.  The properties of the remaining objects were taken
from \citet{ben92}.

\subsubsection{dE Galaxies in the Fundamental Plane}\label{fp_gal}

\begin{figure}
\vskip 0.75 cm
\plotone{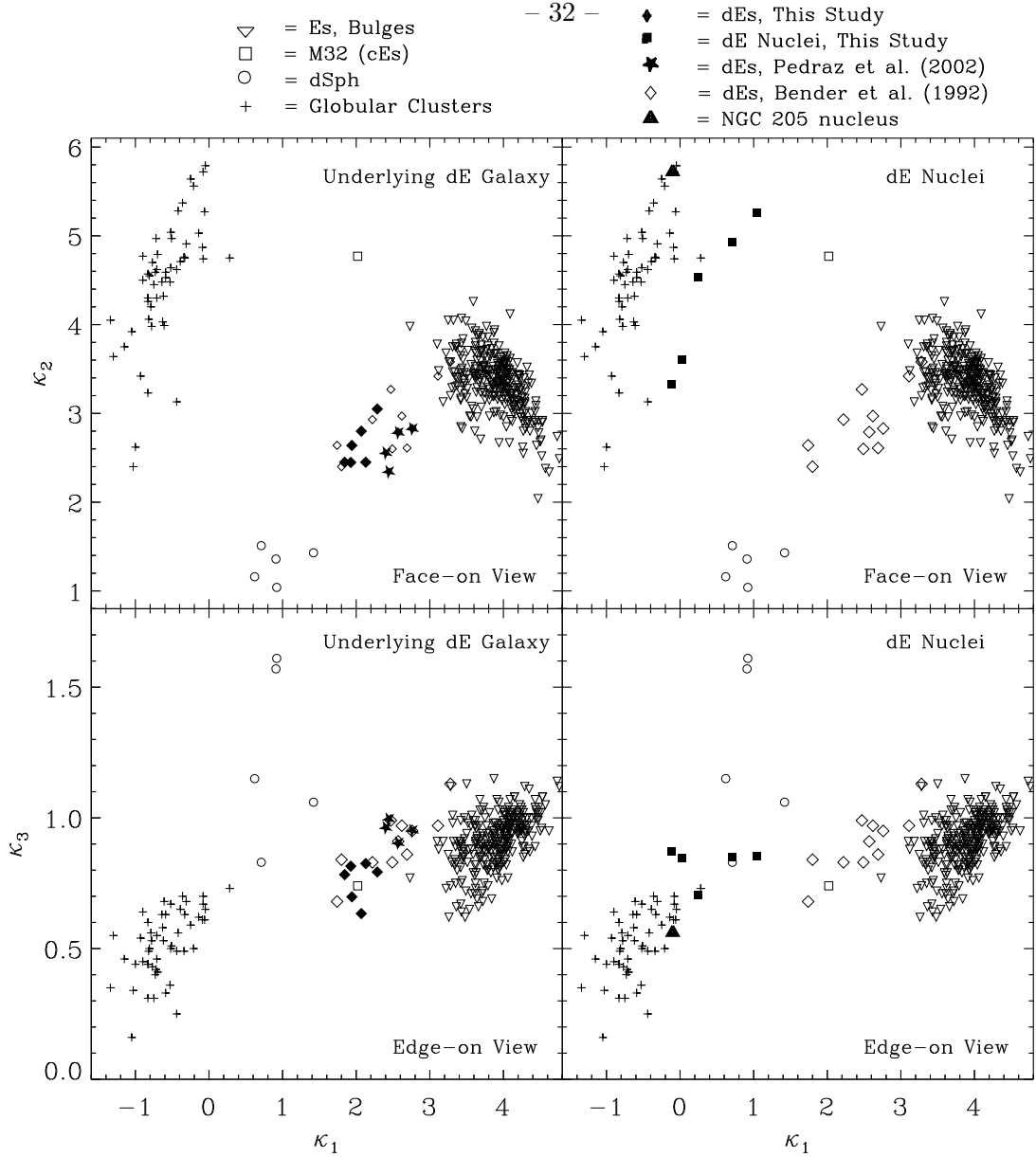}
\vskip 0.6 cm
\caption{Face-on and edge-on projections of the Fundamental Plane for
dynamically hot stellar systems ({\it upper and lower panels, respectively\/}),
where $\kappa_1$, $\kappa_2$, $\kappa_3$ are related to galaxy mass, surface
brightness, and mass-to-light ratio, respectively.  Solid diamonds ({\it left
panels\/}) represent the underlying dE galaxy (minus the nucleus) for our six
Virgo targets; our non-rotating dEs have somewhat smaller masses and $M/L$
ratios than those with substantial rotation studied by \citet{ped02} (solid
stars).  The five observed dE nuclei are plotted as solid squares in the
right panels as they would appear outside their host galaxy; they are
compared to the nucleus of the Local Group dE NGC~205 (solid triangle).  Data
for other systems from \citet{bur97} are (repeated in left and right panels):
classical ellipticals and spiral bulges (open triangles), the compact
elliptical M32 (open square), Local Group and other dEs (open diamonds),
Local Group dwarf spheroidals (open circles), and Galactic globular clusters
(crosses).  The dE nuclei appear to be similar to luminous globular
clusters.\label{fp_fig}}
\end{figure}

As seen in the edge-on, $\kappa_1$ vs.~$\kappa_3$, view of the
Fundamental Plane (lower left panel, Fig.~\ref{fp_fig}) dwarf
ellipticals appear to lie in a plane parallel to, but offset from
classical ellipticals.  In the face-on, $\kappa_1$ vs.~$\kappa_2$,
view (upper left panel, Fig.~\ref{fp_fig}), dEs lie in a very different
region of this plane, on a sequence perpendicular to the locus of
classical ellipticals.  The offset in $\kappa_3$ was first noted by
\citet{ben92} and can be interpreted as either non-homology between
dwarf and classical ellipticals or as a difference in mass-to-light
ratios.  We have shown in Figure~\ref{fig_ml} that dEs tend to have
larger mass-to-light ratios at a given absolute magnitude compared to
classical ellipticals, favoring the latter interpretation of the
$\kappa_3$ offset.  Comparing our non-rotating dEs to the rotating dEs
of \citet{ped02}, these two groups lie in slightly different regions
of the Fundamental Plane.  The rotating dEs lie at larger $\kappa_1$
and $\kappa_3$ than the non-rotating sample, suggesting that they have
both higher masses and mass-to-light ratios.  From the location of the
rotating dEs in the Fundamental Plane, these galaxies are not part of
the low luminosity extension of classical ellipticals known to be
rotationally-supported \citep{dav83}.  Thus, dEs appear to have a
wider range of rotational properties than previously assumed.  The
separation in both luminosity and Fundamental Plane space between
these two samples suggests a correlation between rotation and another
physical quantity, possibly mass.  However, more data is required to
establish such a correlation.

\subsubsection{dE Nuclei in the Fundamental Plane}\label{fp_nuc}

In the right panels of Figure~\ref{fp_fig}, the nuclei of the five
observed nucleated dEs are plotted.  Unlike the underlying dE galaxies
in the Fundamental Plane, dE nuclei lie nearest to the region occupied
by globular clusters.  These $\kappa$-space parameters were calculated
using central velocity dispersions determined directly from a Plummer
model fit to the nucleus.  This is a more accurate estimate of the
nuclear velocity dispersion than the measured central velocity
dispersion, but requires the assumption that the nuclear mass-to-light
ratio equals that of the galaxy, as discussed in
\S\,\ref{nuc_vs_gal_sec}.  This assumption most strongly affects
values of $\kappa_3$.  However, for any reasonable assumed nuclear
mass-to-light ratio, the dE nuclei lie closest to globular clusters in
the all three $\kappa$ indices.  The absolute luminosities of these dE
nuclei ($-8.5 \le M_{\rm V} \le -12.3$) are as bright or brighter than
the most luminous Galactic globular clusters \citep{djo93,har96}.  The
nuclear effective radii are also larger than an average Galactic
globular cluster, but are smaller than the largest known globulars.
The resulting central luminosity densities of dE nuclei, determined
from the Plummer models, are comparable to the average globular
cluster central luminosity density.  The position of the best studied
dE nucleus, that of the Local Group dE NGC~205, a well-resolved
supermassive star cluster of intermediate age and absolute magnitude
$M_V=-9.6$ \citep{jon96}, is also plotted on the Fundamental
Plane.  This nucleus lies squarely in the region occupied by globular
clusters.  The offset position of the Virgo dE nuclei, particularly in
the face-on view of the Fundamental Plane (top right panel
Fig.~\ref{fp_fig}), relative to Galactic globular clusters is most
likely due to larger nuclear masses.

\section{Discussion and Conclusions}\label{disc_sec}

Velocity and velocity dispersion profiles are presented for the major
axes of six dE galaxies in the Virgo Cluster.  These galaxies do not
show evidence for substantial rotation; upper limits on rotation
velocities are well below that expected if these objects were
rotationally flattened.  Dynamical models for these galaxies suggest
mass-to-light ratios in the range $3\le \Upsilon_V \le6$.  We argue
that such ratios are expected for intermediate to old stellar
populations and thus these dEs do not require significant dark matter
inside an effective radius.  Our observations do not rule out
significant dark matter in dEs at larger radii as demonstrated by
giant elliptical galaxies which exist in massive dark halos, but are
not necessarily dark matter dominated at small radii \citep{ger01}.
In Fundamental Plane space, we find that the Virgo dE galaxies,
similar to previously observed dEs, lie in a plane parallel to, but
offset from, that occupied by normal elliptical galaxies.  In this
space, dE nuclei lie near the region occupied by Galactic globular
clusters.

The origin of nuclei in dE galaxies remains an open question.  In the
present sample, there is no obvious difference between the single
non-nucleated dE galaxy (VCC~917) and the underlying galaxies of the
observed nucleated dEs.  The mass-to-light ratio, anisotropy, and
photometric parameters measured for VCC~917 are indistinguishable from
those determined outside the nucleus of the other five dEs.  However,
as a population, non-nucleated dE galaxies in the Virgo Cluster do
have different properties.  They are observed to be less spatially
concentrated, have lower specific globular cluster frequencies, and,
on average, have flatter shapes as compared to nucleated dE galaxies
\citep{san85,mil98,ryd99}.  Proposed scenarios for the origin of dE
nuclei include the remnant cores of larger stripped galaxies
\citep{ger83}, the results of gas infall and star formation or the
coalescence of several globular clusters whose orbits have decayed to
the dE center \citep{oh00}.  We have shown that the observed dE nuclei
share many properties with globular clusters, suggesting similar
formation processes.

Since dE galaxies are preferentially found in dense environments, it
is likely that galaxy interactions play a large role in their
formation and evolution.  The models of \citet{moo98} suggest that
galaxy harassment in clusters can morphologically transform a spiral
galaxy into a dwarf elliptical.  Harassment tends to increase internal
velocity dispersions, but is less efficient in disrupting rotational
motion and is not obviously reconciled with the low rotational
velocities observed in the present dE sample.  If dEs are the
morphologically-transformed remnants of larger progenitor galaxies, a
constraint on such a progenitor population is provided by the central
black hole mass limits determined in \S\,\ref{bh_sec}.  The upper
limit of $\sim10^7\Msun$ for the observed dE galaxies implies that any
dE progenitor must have had a bulge dispersion less then 100\kms,
assuming the $M_{\rm BH}-\sigma_e$ relation \citep{tre02}.  Although
this is not a stringent constraint on dE galaxy formation models,
higher spatial resolution kinematics, and therefore more stringent mass
limits, could be a significant constraint on such models.

As the number of dE galaxies with measured internal kinematics
increases, their position in the Fundamental Plane strengthens the
conclusion that dwarf and classical elliptical galaxies evolve via
very different physical processes.  To determine whether dwarf
ellipticals as a galaxy class evolve under homogeneous conditions
requires more observations.  A critical question is understanding the
apparent dichotomy between the anisotropy-supported dEs presented in
this paper and the rotationally-supported dEs presented by
\citet{ped02} and \citet{der01}.  The fact that rotating and
non-rotating dEs appear to form a ``sequence'' in Fundamental Plane
space, with the latter having somewhat lower mean luminosity, mass,
and mass-to-light ratio, suggests that these are not two distinct
types of dE galaxies but rather are part of a continuous family.
Larger samples are required to establish what, if any, physical
property correlates with the observed rotational velocities and what
this implies for dE galaxy formation.

\acknowledgments

We would like to thank Dennis Zaritsky, Ruth Peterson and Doug Lin for
help with the Keck/HIRES data discussed in
\S\,\ref{esi_hires_comp_sec}.  We are grateful to Bryan Miller for
making his reduced {\it HST\/} WFPC2 images available to us and to
Soeren Larsen for help with the ISHAPE software.  M.G.\ acknowledges
support from the STScI Director's Discretionary Research Fund.




\clearpage

\begin{deluxetable}{lccccccc}
\tabletypesize{\scriptsize}
\tablecaption{Observed Virgo Dwarf Elliptical Galaxies: Basic Properties} 
\tablewidth{0pt}
\tablehead{
\colhead{Galaxy} &
\colhead{$\alpha$ (J2000)} &
\colhead{$\delta$ (J2000)} &
\colhead{Type} &
\colhead{$\epsilon$} &
\colhead{$m_V$} &
\colhead{$A_V$} &
\colhead{$M_{V,0}$}\\
\colhead{}&
\colhead{(h$\,$:$\,$m$\,$:$\,$s)} &
\colhead{($^\circ\,$:$\,'\,$:$\,''$)} &
\colhead{} &
\colhead{} &
\colhead{}&
\colhead{}&
\colhead{}
}
\startdata
VCC~452         & 12:21:04.7 & 11:45:18 & dE4,N & 0.08 & 15.34 & 0.09
& $-15.67$\\
VCC~917/IC~3344 & 12:26:32.4 & 13:34:43 & dE6   & 0.45 & 13.93 & 0.11
& $-17.10$\\
VCC~1073/IC~794 & 12:28:08.6 & 12:05:36 & dE3,N & 0.30 & 13.82 & 0.09
& $-17.19$\\
VCC~1254        & 12:30:05.3 & 08:04:29 & dE0,N & 0.05 & 14.58 & 0.07
& $-16.41$\\
VCC~1577        & 12:34:38.4 & 15:36:10 & dE4,N & 0.17 & 15.24 & 0.09
& $-15.77$\\
VCC~1876/IC~3658& 12:41:20.4 & 14:42:02 & dE5,N & 0.23 & 14.65 & 0.10
& $-16.37$\\
\enddata
\tablecomments{Galaxy classifications taken from \citet{bin85}.  The
ellipticity $\epsilon$ is measured between $1''<r<20''$; apparent and
absolute magnitudes are determined inside an $r<40''$ aperture.  The
absolute magnitudes, $M_{V,0}$, assume a Virgo Cluster distance modulus
of $(m - M)_0 = 30.92$ and are corrected for reddening.}
\end{deluxetable} 

\vskip 1cm
\begin{deluxetable}{lcrc}
\tabletypesize{\scriptsize}
\tablecaption{Spectroscopic Observing Parameters}
\tablewidth{0pt}
\tablehead{
\colhead{Galaxy} &
\colhead{Exposure Time} &
\colhead{PA$_{\rm maj}$}&
\colhead{Seeing FWHM} \\
\colhead{}&
\colhead{(s)}&
\colhead{($^{\circ}$)} & 
\colhead{($''$)} 
}
\startdata
VCC~452         & 3600 & $-34$~~~ & 1.1\\
VCC~917/IC~3344 & 3600 &  52~~~   & 0.9\\
VCC~1073/IC~794 & 3600 & $-60$~~~ & 0.8\\
VCC~1254        & 3600 &   0~~~   & 0.8\\
VCC~1577        & 6000 &  20~~~   & 1.0\\ 
VCC~1876/IC~3658& 3600 &  68~~~   & 1.0\\ 
\enddata
\end{deluxetable}

\begin{deluxetable}{lccccccc}
\tabletypesize{\scriptsize}
\tablecaption{Surface Photometry Results} 
\tablewidth{0pt}
\tablehead{
\colhead{Galaxy} &
\colhead{$r_{\rm eff}$} &
\colhead{$\mu_{V,\rm eff}$}&
\colhead{$n_{\rm Sersic}$}&
\colhead{$b_{\rm nuc}$}&
\colhead{$\mu_{V,\rm eff}^{\rm nuc}$}&
\colhead{$m_V^{\rm nuc}$}&
\colhead{$M_{V,0}^{\rm nuc}$}\\
\colhead{}&
\colhead{[ $''$ (kpc)]}&
\colhead{(mag arcsec$^{-2}$)}&
\colhead{}&
\colhead{[ $''$ (kpc)]}&
\colhead{(mag arcsec$^{-2}$)}&
\colhead{}&
\colhead{}
}
\startdata
VCC~452         &$\>$~9.6 (0.71)& 22.3 & 1.6 & 0.15 (0.011) & 20.7 & 22.54
&~$\,-8.47$\\
VCC~917/IC~3344 & 12.2 (0.90)   & 21.4 & 2.9 &     ...      & ...  & ...
&  ~...    \\ 
VCC~1073/IC~794 & 11.1 (0.82)   & 21.1 & 1.9 & 0.13 (0.010) & 17.4 & 19.86
&  $-11.15$\\
VCC~1254        & 14.4 (1.07)   & 22.4 & 2.9 & 0.17 (0.013) & 16.7 & 18.67
&  $-12.32$\\
VCC~1577        & 10.5 (0.78)   & 22.4 & 1.1 & 0.16 (0.011) & 20.1 & 22.23
&~$\,-8.78$\\
VCC~1876/IC~3658& 10.5 (0.78)   & 21.8 & 0.8 & 0.11 (0.008) & 17.9 & 20.92
&  $-10.10$\\
\enddata
\tablecomments{The effective half-light radius $r_{\rm eff}$, effective
surface brightness $\mu_{V,\rm eff}$, and index $n_{\rm Sersic}$ are
determined by fitting a Sersic model to the galaxy surface brightness profile
outside $r>1''$.  The effective half-light radius $b_{\rm nuc}$, effective
surface brightness $\mu_{V,\rm eff}^{\rm nuc}$, and total magnitude of the
nucleus are determined only for the nucleated dEs by fitting a Plummer
model to the inner ($r<1''$) brightness profile via deconvolution using the
ISHAPE software \citep{lar99} as explained in \S\,\ref{sbfits}.}
\end{deluxetable}

\begin{deluxetable}{lcccccc}
\tabletypesize{\scriptsize}
\tablecaption{Results of Kinematical Analysis / Dynamical Modeling} 
\tablewidth{0pt}
\tablehead{
\colhead{Galaxy} &
\colhead{$v_{\rm sys}$} &
\colhead{$v_{\rm rot}$}&
\colhead{$\sigma_{\rm gal}$} & 
\colhead{$\sigma_{\rm nuc}$} & 
\colhead{$\Upsilon_V$}&
\colhead{$\beta$}\\
\colhead{}&
\colhead{(\kms)}&
\colhead{(\kms)}&
\colhead{(\kms)}&
\colhead{(\kms)}&
\colhead{[$(\Upsilon_V)_\odot$]}&
\colhead{}
}
\startdata
VCC~452         &    1380& $1.0\pm1.7$ & 23.8 &$\,$~7.7& $5.28\pm0.82$ &   $-
0.27\pm0.23$\\
VCC~917/IC~3344 &    1186& $0.4\pm0.4$ & 31.1 &    ... & $3.41\pm0.14$ &   $-
0.87\pm0.15$\\
VCC~1073/IC~794 &    1862& $2.1\pm0.4$ & 44.6 &    31.7& $5.83\pm0.17$ &   $-
0.64\pm0.08$\\
VCC~1254        &    1220& $0.9\pm0.9$ & 31.0 &    48.6& $5.99\pm0.17$ &   $-
2.56\pm0.98$\\
VCC~1577        &$\,$~361& $1.3\pm0.7$ & 26.8 &$\,$~9.8& $5.76\pm0.86$ &   $-
0.07\pm0.17$\\ 
VCC~1876/IC~3658&$\,$~~95& $1.2\pm1.4$ & 25.7 &    16.5& $3.33\pm0.83$ 
&$\>~~0.10\pm0.23$\\
\enddata
\tablecomments{The heliocentric systemic velocity $v_{\rm sys}$ is determined
from the mean of each velocity profile.  The rotation speed $v_{\rm rot}$
listed in most cases is an upper limit; $\sigma_{\rm gal}$ is the average
line-of-sight velocity dispersion outside $r>1''$; $\sigma_{\rm nuc}$ is the
calculated projected velocity dispersion of a Plummer profile fit to the dE
nucleus, assuming it has the same mass-to-light ratio ($\Upsilon_V$) as the
underlying galaxy; $\Upsilon_V$, the anisotropy parameter $\beta$, and their
corresponding errors are determined from the dynamical modeling discussed in
\S\,\ref{ml_aniso_sec}.}
\end{deluxetable}

\begin{deluxetable}{lcccccc}
\tabletypesize{\scriptsize}
\tablecaption{Fundamental Plane $\kappa$-Space Parameters}
\tablewidth{0pt}
\tablehead{
\colhead{Galaxy} &
\multicolumn{3}{c}{Underlying dE Galaxy} &
\multicolumn{3}{c}{dE Nuclei} \\
\colhead{} & 
\colhead{$\kappa_1$} & 
\colhead{$\kappa_2$} & 
\colhead{$\kappa_3$} & 
\colhead{$\kappa_1$} & 
\colhead{$\kappa_2$} & 
\colhead{$\kappa_3$} 
}
\startdata
VCC~452         & 1.84 & 2.45 & 0.78 & $-0.11$ & 3.33 & 0.87 \\
VCC~917/IC~3344 & 2.07 & 2.80 & 0.63 &  ~...   & ...  & ...  \\
VCC~1073/IC~794 & 2.29 & 3.05 & 0.79 & ~~0.71  & 4.93 & 0.85 \\
VCC~1254        & 2.13 & 2.45 & 0.83 & ~~1.04  & 5.26 & 0.86 \\
VCC~1577        & 1.93 & 2.45 & 0.82 & ~~0.03  & 3.61 & 0.85 \\
VCC~1876/IC~3658& 1.94 & 2.64 & 0.70 & ~~0.24  & 4.53 & 0.71 \\ 
\enddata
\end{deluxetable}

\clearpage

\begin{thebibliography}{}
\bibitem[Bender et~al.(1992)Bender, Burstein, \& Faber]{ben92} Bender, R.,
Burstein, D., \& Faber, S.~M.\ 1992, \apj, 399, 462

\bibitem[Bender \& Nieto(1990)]{ben90} Bender, R., \& Nieto, J.-L.\ 1990,
A\&A, 239, 97

\bibitem[Bender et~al.(1991)Bender, Paquet, \& Nieto]{ben91} Bender, R.,
Paquet, A., \& Nieto, J.-L.\ 1991, A\&A, 246, 349

\bibitem[Bingelli \& Jerjen(1998)]{bin98} Bingelli, B., \& Jerjen, H.\ 1998,
A\&A, 333, 17

\bibitem[Bingelli et~al.(1985)Bingelli, Sandage, \& Tammann]{bin85} Bingelli,
B., Sandage, A., \& Tammann, G.~A.\ 1985, \aj, 90, 1681

\bibitem[Bingelli et~al.(1987)Bingelli, Tammann, \& Sandage]{bts87} Bingelli,
B., Tammann, G.~A., \& Sandage, A.\ 1987, \aj, 94, 251

\bibitem[Binney \& Tremaine(1987)]{bin87} Binney, J., \& Tremaine, S.\
1987, Galactic Dynamics, Princeton University Press, Princeton

\bibitem[Burstein et~al.(1997)]{bur97} Burstein, D., Bender, R., Faber,
S.~M., \& Nolthenius, R.\ 1997, \aj, 114, 1365

\bibitem[Davies et~al.(1983)]{dav83} Davies, R.~L., Efstathiou, G.,
Fall, S.~M., Illingworth, G., \& Schechter, P.~L.\ 1983, \apj, 266, 41

\bibitem[Dejonghe(1987)]{dej87} Dejonghe, H.\ 1987, \mnras, 224, 13

\bibitem[Dekel \& Silk(1986)]{dek86} Dekel, A., \& Silk, J.\ 1986, \apj, 303,
39

\bibitem[de Vaucouleurs(1948)]{dev48} de Vaucouleurs, G.\ 1948,
Ann. d'Astrophys., 11, 247

\bibitem[Djorgovski(1993)]{djo93} Djorgovski, S.~G.\ 1993, in
``Structure and Dynamics of Globular Clusters'', ed.\ S.~G.~Djorgovski
and G.~Meylan (San Fransisco: ASP), 373

\bibitem[Durell(1997)]{dur97} Durell, P.~R.\ 1997, AJ, 113, 531

\bibitem[Ferrarese \& Merritt(2000)]{fer00} Ferrarese, L., \& Merritt,
D.\ 2000, \apj, 539, L9

\bibitem[Ferguson \& Binggeli(1994)]{fer94} Ferguson, H.~C., \& Binggeli, B.\
1994, A\&A Rev., 6, 67

\bibitem[Freedman et~al.(2001)]{fre01} Freedman, W.~L., Madore, B.~F.,
Gibson, B.~K., Ferrarese, L., Kelson, D.~D., Sakai, S., Mould, J.~R.,
Kennicutt, R.~C., Jr., Ford, H.~C., Graham, J.~A., Huchra, J.~P., Hughes,
S.~M.~G., Illingworth, G.~D., Macri, L.~M., \& Stetson, P.~B.\ 2001, \apj,
553, 47

\bibitem[Gavazzi et~al.(2001)]{gav01} Gavazzi, G., Zibetti, S.,
Boselli, A., Franzetti, P., Scodeggio, M., \& Martocchi, S.\ 2001, A\&A,
372, 29

\bibitem[Gebhardt et~al.(2000)]{geb00} Gebhardt, K., Bender, R., Bower,
G., Dressler, A., Faber, S.~M., Filippenko, A.~V., Green, R.,
Grillmair, C., Ho, L.~C., Kormendy, J., Lauer, T., Magorrian, J.,
Pinkney, J., Richstone, D., \& Tremaine, S.\ 2000, \apj, 539, L13

\bibitem[Geha et~al.(2002)Geha, Guhathakurta, \& van der Marel]{geh02} Geha,
M., Guhathakurta P., \& van der Marel, R.~P.\ 2002, in ``The Shapes of
Galaxies and their Halos'', ed.\ P.~Natarajan, World Scientific, in press
(astro-ph/0107010)

\bibitem[Gerhard et~al.(2001)]{ger01} Gerhard, O., Kronawitter, A.,
Saglia, R.~P., \& Bender, R.\ 2001, \aj, 121,1936

\bibitem[Gerola et~al.(1983)]{ger83} Gerola, H., Carnevali, P., \&
Salpeter, E.~E.\ 1983, \apj, 268, L75

\bibitem[Harris(1996)]{har96} Harris, W.~E.\ 1996, \aj, 112, 1487

\bibitem[Held et~al.(1990)Held, Mould, \& de~Zeeuw]{hel90} Held, E.~V.,
Mould, J.~R., \& de~Zeeuw, P.~T.\ 1990, \aj, 100, 415

\bibitem[Holtzman et~al.(1995)]{hol95} Holtzman, J.~A., Burrows, C.~J.,
Casertano, S., Hester, J.~J., Trauger, J.~T., Watson, A.~M., \& Worthey, G.\
1995, \pasp, 107, 1065

\bibitem[Jones et~al.(1996)]{jon96} Jones, D.~H., Mould, J.~R.,
Watson, A.~M., Grillmair, C., Gallagher, J.~S., Ballester, G.~E.,
Burrows, C.~J., Casertano, S., Clarke, J.~T., Crisp, D., Griffiths,
R.~E., Hester, J.~J., Hoessel, J.~G., Holtzman, J.~A., Scowen, P.,
Stapelfeldt, K.~R., Trauger, J.~T., \& Westphal, J.~A.\ 1996, \apj,
466, 742

\bibitem[Krist \& Hook(1997)]{kri97} Krist, J., \& Hook, R.\ 1997, The
TinyTim User's Guide (Baltimore: STScI)

\bibitem[Larsen(1999)]{lar99} Larsen, S.~S. 1999, A\&A Suppl., 139, 393

\bibitem[Lauer et~al.(1995)]{lau95} Lauer, T.~R., Ajhar, E.~A., Byun, Y.-I.,
Dressler, A., Faber, S.~M., Grillmair C., Kormendy, J., Richstone, D., \&
Tremaine, S.\ 1995, \aj, 110, 2622

\bibitem[Magorrian et~al.(1998)]{mag98} Magorrian, J., Tremaine, S.,
Richstone, D., Bender, R., Bower, G., Dressler, A., Faber, S.~M.,
Gebhardt, K., Green, R., Grillmair, C., Kormendy, J., \& Lauer, T.\ 1998,
\aj, 115, 2285

\bibitem[Mayer et~al.(2001)]{may01} Mayer, L., Governato, F., Colpi, M.,
Moore, B., Quinn, T., Wadsley, J., Stadel, J., \& Lake, G.\ 2001, \apj, 559,
754

\bibitem[Miller et~al.(1998)]{mil98} Miller, B.~W., Lotz, J.~M., Ferguson,
H.~C., Stiavelli, M., \& Whitmore, B.~C.\ 1998, \apj, 508, L133

\bibitem[Moore et~al.(1998)Moore, Lake, \& Katz]{moo98} Moore, B., Lake, G.,
\& Katz, N.\ 1998, \apj, 495, 139

\bibitem[Oh \& Lin(2000)]{oh00} Oh, K.~S., \& Lin, D.~N.~C.\ 2000, \apj, 543,
620

\bibitem[Pedraz et al.(2002)]{ped02} Pedraz, S., Gorgas, J., Cardiel,
N., Sanchez-Blazquez, P., \& Guzman, R.\ 2002, \mnras, 332, L59

\bibitem[Peterson \& Caldwell(1993)]{pet93} Peterson, R.~C., \& Caldwell,
N.\ 1993, \aj, 105, 1411

\bibitem[Press et~al.(1992)]{pre92} Press, W.~H., Teukolsky, S.~A.,
Vetterling, W.~T., Flannery, B.~P.\ 1992, Numerical Recipes, Cambridge
University Press, Cambridge, \S 15.6

\bibitem[de Rijcke et~al.(2001)]{der01} de~Rijcke, S., Dejonghe, H.,
Zeilinger, W.~W., \& Hau, G.~K.~T.\ 2001, \apj, 559, L21

\bibitem[Ryden et~al.(1999)]{ryd99} Ryden, B., Terndrup, D.~M., Pogge, R.~W.,
\& Lauer, T.~R.\ 1999, \apj, 517, 650

\bibitem[Sandage, Bingelli, \& Tammann(1985)]{san85} Sandage,
A., Binggeli, B., \& Tammann, G.\ 1985, \aj, 90, 1759

\bibitem[Schlegel et~al.(1998)Schlegel, Finkbeiner, \& Davis]{sch98}
Schlegel, D.~J., Finkbeiner, D.~P., \& Davis, M.\ 1998, \apj, 500, 525

\bibitem[S\'{e}rsic(1968)]{ser68} S\'{e}rsic, J.~L.\ 1968, Atlas de Galaxias
Australes (C\'{o}rdoba: Obs. Astron., Univ. Nac. C\'{o}rdoba)

\bibitem[Sheinis et~al.(2002)]{she02} Sheinis, A.~I., Bolte, M., Epps,
H.~W., Kibrick, R.~I., Miller, J.~S., Radovan M.~V., Bigelow, B.~C., \&
Sutin, B.~M.\ 2002, \pasp, 114, 851

\bibitem[Stiavelli et~al.(2001)] {sti01} Stiavelli, M., Miller, B.~W.,
Ferguson, H.~C., Mack, J., Whitmore, B.~C., \& Lotz, J.~M.\ 2001, \aj, 121,
1385

\bibitem[Tremaine et~al.(2002)]{tre02} Tremaine, S., Gebhardt, K.,
Bender, R., Bower, G., Dressler, A., Faber, S.~M., Filippenko, A.~V.,
Green, R., Grillmair, C., Ho, L.~C., Kormendy, J., Lauer, T.~R.,
Magorrian, J., Pinkney, J., \& Richstone, D.\ 2002, \apj, 574, 740

\bibitem[van der Marel(1994)]{vdm94} van der Marel, R.~P.\ 1994, \mnras, 270,
271 

\bibitem[van der Marel \& Franx(1993)]{vdm93} van der Marel, R.~P., \&
Franx, M.\ 1993, \apj, 407, 525

\bibitem[Worthey(1994)]{wor94} Worthey, G.\ 1994, \apjs, 95, 107

\end{thebibliography}
\end{document}